\documentclass[a4paper,fleqn,usenatbib]{mnras}
\usepackage[T1]{fontenc}
\usepackage{ae,aecompl}

\usepackage{graphicx}	% Including figure files
\usepackage{amsmath}	% Advanced maths commands
\usepackage{amssymb}	% Extra maths symbols

\newcommand{\be}{\begin{equation}}
\newcommand{\ee}{\end{equation}}
\newcommand{\bea}{\begin{eqnarray}}
\newcommand{\eea}{\end{eqnarray}}

\title[Red Discs and Blue Bulges]{NoSOCS in SDSS. V. Red Disc and Blue Bulge Galaxies Across Different Environments}
\author[Lopes et al.]{P. A. A. Lopes$^{1}$\thanks{E-mail: 
    plopes@astro.ufrj.br}, S. B. Rembold$^{2}$, A. L. B. Ribeiro$^{3}$,
  R. S. Nascimento$^{1}$, B. Vajgel$^{4}$\\
\\
$^{1}$Observat\'orio do Valongo, Universidade Federal do Rio de Janeiro, 
Ladeira do Pedro Ant\^onio 43,\\
Rio de Janeiro, RJ, 20080-090, Brazil\\
$^{2}$Universidade Federal de Santa Maria -- 97105-900, Santa Maria-RS, Brazil\\
$^{3}$Laborat\'orio de Astrof\'isica Te\'orica e Observacional -- Departamento 
de Ci\^encias Exatas e Tecnol\'ogicas -- \\
Universidade Estadual de Santa Cruz, 45650-000, Ilh\'eus, BA, Brazil\\
$^{4}$ Observat\'orio Nacional, Rua General Jos\'e Cristino, 77,
Rio de Janeiro, RJ, 20921-400, Brazil\\
}
\date{Accepted 2016 June 20. Received 2016 May 23; in original form 2016 March 12}

\pubyear{2016}

\begin{document}
\label{firstpage}
\pagerange{\pageref{firstpage}--\pageref{lastpage}}
\maketitle

\begin{abstract}
  We investigated the typical environment and physical properties of
  ``red discs'' and ``blue bulges'', comparing those to the ``normal''
  objects in the blue cloud and red sequence. Our sample is composed
  of cluster members and field galaxies at $z \le 0.1$, so that we can assess
  the impact of the local and global environment. We find that disc galaxies
  display a strong dependence on environment, becoming redder for higher
  densities. This effect is more pronounced for objects within the virial
  radius, being also strong related to the stellar
  mass. We find that local and global environment affect galaxy properties,
  but the most effective parameter is stellar mass. We find evidence
  for a scenario where ``blue discs'' are transformed into ``red discs'' as
  they grow in mass and move to the inner parts of clusters. From the
  metallicity differences of red and blue discs, and the analysis of their
  star formation histories, we suggest the quenching process is slow. We
  estimate a quenching time scale of $\sim $ 2$-$3 Gyr. We also find from
  the sSFR$-$M$_*$ plane that ``red discs'' gradually change as they move
  into clusters. The ``blue bulges'' have many similar properties than
  ``blue discs'', but some of the former show strong signs of asymmetry.
  The high asymmetry ``blue bulges'' display enhanced recent star formation
  compared to their regular counterparts. That indicates some of these
  systems may have increased their star formation due to mergers.
  Nonetheless, there may not be a single evolutionary path for these
  blue early-type objects.
\end{abstract}


%%%%%%%%%%%%%%%%%%%%%%%%%%%%%%%%%%%%%%%%%%%%%%%%%%

%%%%%%%%%%%%%%%%% BODY OF PAPER %%%%%%%%%%%%%%%%%%

\section{Introduction}

The environment where galaxies reside have a strong influence on their
physical properties \citep{oem74, dre80}, related to
their structure and star formation activity.
In a color-magnitude (or color-mass) diagram we can broadly classify local
galaxies in two main types. In such diagrams most galaxies are found in two
regions, called ``blue cloud'' (BC) and ``red sequence'' (RS). The former is
mainly composed by late-type galaxies, blue spirals, with high star formation
rate (SFR). The latter comprises early-types, elliptical and S0 galaxies,
which are redder, bulge-dominated and with negligible star formation. Galaxies
are believed to migrate from the BC to the RS passing through a region with
low number density of galaxies, known as the ``Green Valley'' (GV,
\citealt{wyd07}). GV galaxies are considered a transitory population
between the BC and RS \citep{sal09}. Galaxies in this region define this
low number density ``valley'' because they are undergoing a rapid evolution
phase, when their star-formation is being quenched \citep{cra14}.

The subject of transitional galaxy types has already been investigated for a
few decades. For instance, some well known transitional objects are the
post-starburst (``E+A'') galaxies \citep{dre83, zab96, tra03, got05, swi12}.
and ultra-luminous infrared galaxies \citep{soi84, cap06, cap09}.
More recently, several studies in the literature
aim to investigate the properties and evolution of different  transitional
populations. \citet{wol09} investigated the properties
of optically passive spirals in the cluster A901/2;
\citet{cro14} investigated RS galaxies with residual star formation
in massive nearby clusters; \citet{mci14} studied
recently quenched elliptical galaxies (RQEs) in the local Universe;
\citet{sme15} present evidence of different star formation
histories of GV objects; and \citet{cra14} investigate the nature
of GV and luminous compact blue galaxies (LCBGs) and their relation to
BC and RS objects.

Galaxy stellar population properties also show a strong dependence
on galaxy stellar mass \citep{tor10, sal09, gug15}.
On average, high mass galaxies formed most of their
stars at earlier epochs and in a much shorter time scale than lower mass
systems. That effect is called downsizing \citep{cow96, bun06, cim06, fon09}.
Disentangling the importance of environment and stellar mass, and even
galaxy morphology, is a hard task. Currently, there is still a debate if the
environmental dependence could simply be the result of different mass and/or
morphological distributions with environment. In general, different studies
indicate that both galaxy mass and environment are important. At least on
what regards quenching, the environment is more relevant for lower mass
objects \citep{hai07, vul15}. In a recent study \citep{gug15}
verify that current morphology is correlated to the
star formation activity, but is not important for the stellar history. They
also find the average star formation history (SFH) depends on galaxy mass,
but at fixed mass the SFH depends on the environment.

This work is the fifth of a series aiming to investigate cluster and
galaxies' properties at low redshifts ($z \le 0.1$). Our main goal is 
to study transitional galaxy populations characterizing their main
properties in different environments, from the extreme field to the
central parts of groups/clusters. For that purpose we separate our galaxy
sample in four galaxy populations, two representing the ``normal'' galaxy
bimodal distribution found in the local Universe, and two associated to
intermediate types or galaxies in transition. In a similar way to what
has been done by \citet{cra14} we simply consider photometric
derived parameters for that separation. In our case, we use the (u-r) color
and the concentration index ($C = R_{90}/R_{50}$). According to these
parameters we separate the sample, calling the ``normal'' galaxies as
``red bulges'' and ``blue discs'', and the transitional types as
``red discs'' and ``blue bulges''.  We compare several physical properties
(such as age, metallicity, SFR, and stellar mass) of these four classes
below, as well as their environmental variation from the field to clusters.
The analysis is based on two complementary luminosity ranges 
($M_r \le M^*+1$ and $M^*+1 < M_r \le M^*+3$), where $M^*$ is the
characteristic magnitude of the luminosity function in the $r-$band.

This paper is organized as follows: $\S$2 has the data description, where we 
also discuss the field sample, the local galaxy density estimates, the
stellar population properties from different codes, visual
morphologies, and the separation of the galaxy populations.
In $\S$3 we present the environmental variation of the transitional galaxy
populations. In $\S$4, to better characterize the samples, we compare physical
properties and environmental parameters of the four galaxy populations.
In $\S$5 we try to disentangle the importance of stellar mass and environment
(local and global) for the galaxy populations. We also try to understand the
nature of the transitional populations from their inspection in the
$Z-$M$_*$, Age$-$M$_*$ and sSFR$-$M$_*$ planes, and the analysis of their
star formation histories (SFH). In $\S$6 we have a discussion on the quenching
time of the ``blue disc'' population and on the nature of the transitional
objects. We summarize our main results in $\S$7. 
The cosmology assumed in this work considers $\Omega_{\rm m}=$0.3, 
$\Omega_{\lambda}=$0.7, and H$_0 = 100$ $\rm h$ $\rm km$ $s^{-1}$ Mpc$^{-1}$, 
with $\rm h$ set to 0.7. For simplicity, in the following we are going to use 
the term ``cluster'' to refer loosely to groups and clusters of galaxies.

\section{Data}

This work is based on photometric and spectroscopic data from the
SDSS, as well as Wide-field Infrared Survey Explorer (WISE) photometry.
In what follows we describe the
cluster and field samples, and the main properties of the galaxy
data we use.

\subsection {Cluster sample}

In the first paper of this series (hereafter paper I, \citealt{lop09a}) 
we defined a cluster sample from the supplemental version of the Northern Sky 
Optical Cluster Survey (NoSOCS, \citealt{lop04}). For that we used data
from the 5th Sloan Digital Sky Survey (SDSS) release, from which we
re-estimated photometric redshifts as in \citet{lop07}. This sample comprises 
7,414 systems well sampled in SDSS DR5 (details in paper I). NoSOCS has its origin 
on the digitized version of the Second Palomar Observatory Sky Survey (POSS-II; 
DPOSS, \citealt{djo03}). In \citet{gal04} and \citet{ode04} the photometric calibration 
and object classification for DPOSS, respectively, are described. The 
supplemental version of NoSOCS \citep{lop04} goes deeper ($z \sim 0.5$), but 
covers a smaller region than the main NoSOCS catalog \citep{gal03, gal09}.

For a subset of the 7,414 NoSOCS systems with SDSS data we extracted a 
subsample of  low redshift galaxy clusters ($z \le 0.100$). This subsample
comprises 127 clusters, for which we had enough spectra in SDSS for 
spectroscopic redshift determination, as well as to select cluster members and 
perform a virial analysis, obtaining estimates of velocity dispersion, 
physical radius and mass ($\sigma_P$, $R_{500}$, $R_{200}$, $M_{500}$ and 
$M_{200}$; details in paper I). This low-redshift sample was 
complemented with more massive systems
from the Cluster Infall Regions in SDSS (CIRS) sample (\citealt{rin06}).
CIRS is a collection of $z \le 0.100$ X-ray selected clusters 
overlapping the SDSS DR4 footprint. The same cluster parameters listed above 
were determined for these 56 CIRS clusters. 

In the second paper of this 
series (hereafter paper II, \citealt{lop09b}) we investigated the scaling 
relations of clusters using this combined sample of 183 clusters at 
$z \le 0.100$, except for three systems that are not used for having  
biased values of $\sigma_P$ and mass due to projection effects.  
For the clusters with at least five galaxy members within 
R$_{200}$ we also have a substructure estimate, based on the DS, or $\Delta$ 
test \citep{dre88}. Details about this
low-redshift sample and the estimates obtained for the clusters can be
found in papers I and II.

The redshift limit of the sample ($z = 0.100$) is due to incompleteness in
the SDSS spectroscopic survey for higher redshifts, where galaxies fainter 
than $M^*+1$ are missed, biasing the dynamical analysis (see discussion in section 
4.3 of \citealt{lop09a}). We eliminated interlopers using the ``shifting gapper'' 
technique \citep{fad96}, applied to all galaxies with spectra available within a 
maximum aperture of 2.50 h$^{-1}$ Mpc. We also estimated X-ray luminosity ($L_X$, 
using ROSAT All Sky Survey data), optical luminosity ($L_{opt}$) and richness 
(N$_{gals}$, \citealt{lop09a, lop09b}). The centroid of each NoSOCS cluster is a 
luminosity weighted estimate, which correlates well with the X-ray peak 
(see \citealt{lop06}).

In \citet{rib13} (hereafter paper III) we investigated the connection between
galaxy evolution and the dynamical state of galaxy clusters, indicated by their
velocity distributions. In \citet{lop14} (hereafter paper IV) we investigate
the role of environment from the field, through the outer regions of clusters
and their cores. In a forthcoming paper (Rembold, Ribeiro \& Lopes) we will
study the properties of brightest cluster galaxies and their connection to
the parent clusters. Here we focus on the investigation 
of the properties of transitional galaxies (such as metallicity,
L$_{dust}$, and star formation rate), and their relations with
the environment and stellar mass. Note
that for the cluster regions we only use galaxies that are selected
as cluster members by the ``shifting gapper'' technique. A control field
sample is described below.

In \citet{lop14} we implemented 
one modification to the ``shifting gapper'' technique, that resulted in the
exclusion of a few lower mass systems (details in paper IV). Due to that our
final sample comprises 152 groups and clusters, for which we have
6,415 galaxies, being 5,106 with $M_r \le M^*+1$ 
(from clusters at $z \le 0.100$) and 1,309 galaxies with 
$M^*+1 < M_r \le M^*+3$ (from objects at $z \le 0.045$). Our clusters span 
the range $150 \la \sigma_P \la 950$ km s$^{-1}$, or the equivalently in terms 
of mass, $10^{13} \la M_{200} \la 10^{15} M_{\odot}$. 
In that fourth paper we had investigated the role of
environment beyond the extent
of galaxy clusters, corroborating a scenario on which pre-processing in
groups leads to a strong evolution in galaxy properties, before they are
accreted by large clusters. This final galaxy sample, the cluster properties
derived above, the density estimates and the field data, were all based on
the SDSS DR7.

\subsection{Field sample}

The galaxy field sample is constructed as
follows. From the whole DR7 data set we select galaxies that would not
be associated to a group or cluster, considering the cluster catalog
from \citet{gal09}. To be conservative, we consider a galaxy to belong to
the \emph{field} if it is not found within 4.0 Mpc and not having a 
redshift offset smaller than 0.06 of any cluster from \citet{gal09}. We
select more than 60,000 field galaxies, but work with a smaller subset
(randomly chosen) for which we derived local density estimates. 
In the end we use 2,936 field galaxies at $z \le$ 0.100 with
$M_r \le M^*+1$,  and 1,740 at $z \le$ 0.045 with $M^*+1 < M_r \le M^*+3$
(the same order of cluster galaxies). For these objects we compute density
estimates (see below) in the same way as done for the cluster members.

Note this field
sample is based on a comparison to one cluster catalog \citep{gal09}. As any
other cluster catalog, the one from \citet{gal09} is complete for rich
systems, but not for the smaller mass groups and clusters. Even if we were
using all group and cluster catalogs available in the literature we would
still be incomplete for low mass systems. Due to that we may select galaxies
as field objects which may actually belong to small groups. Their local
densities will then be high (see for instance Figs.~\ref{fig:denbins2} and
\ref{fig:denbinsb2} below, or
Figs. 3 and 4 of \citealt{lop14}), with typical values
Log $\Sigma_5 > 0$. However, the number of objects in this case is not
large (as can be seen by the ``scale'' represented by the error bars in
these figures), and we decided to leave them in the field sample, instead
of making an arbitrary cut in $\Sigma_5$. Another reason explaining these
cases is the fact that we consider galaxies out to large distances from the
clusters, so that some cluster galaxies have small local densities
(Log $\Sigma_5 < 0.8$). The density overlap (between field and clusters
results) we see in Figs.~\ref{fig:denbins2} and ~\ref{fig:denbinsb2} happens
for cluster galaxies that are in the infall (far from the center).

\subsection {Local Galaxy Density Estimates}

Considering the results from \citet{mul12} we decided to adopt a 
nearest neighbour method to estimate the local environment.
The authors also mention
that the choice of $n$ - the rank of the density-defining neighbour - is very 
important, as the environment measure may loose power
in the case $n$ is larger than the number of galaxies residing in the halo.
Hence, we chose to work with the $\Sigma_5$ local galaxy density estimator, as
$n = 5$ is typically smaller than the number of galaxies we have per cluster and
is a common estimate in the literature. To estimate local galaxy
densities we proceed as follows. For every galaxy we compute its
projected distance, d$_5$, to the 
5th nearest galaxy found around it, within a maximum 
velocity offset of 1000 $km$ $s^{-1}$ 
(relative to the velocity of the galaxy in question). The  local  density 
$\Sigma_5$  is simply given by 5/$\pi$d$_N^{2}$, and is measured in  units of 
galaxies/Mpc$^2$. Density estimates are also obtained relatively only
to galaxies brighter than a fixed luminosity range, which we adopt
as $M^* + 1.0$. On what regards the global environment we consider
the distance to the center of the parent cluster, the cluster mass, or the
comparison between field and cluster results.

\subsubsection{Correction for the fiber collision issue}

The fiber collision issue affects the SDSS spectroscopic sample and the derived
density estimates. Due to a mechanical restriction spectroscopic fibers cannot  
be placed closer than 55 arcsecs on the sky.  An algorithm used for  target  
selection randomly  chooses  which  galaxy  gets a  fiber, in case of
a conflict \citep{str02}. This  problem is reduced by spectroscopic plate
overlaps, but   fiber  collisions   still   lead   to  a   $\sim$ 6\%
incompleteness in  the main galaxy  sample. Our  approach to  fix
this problem is similar to  the one adopted by \citet{ber06} and 
\citet{lab10}. For galaxies brighter than $r =$ 18 with no redshifts
we  assume the redshift of the nearest neighbour  on the sky (generally
the galaxy it collided with). This may result in some nearby galaxies to
be placed at high redshift, artificially increasing their  estimated
luminosities. Due to that the collided  galaxies  also assume the
magnitudes of their nearest neighbours, resulting in an unbiased 
luminosity distribution. Notice that the fraction of fixed galaxies is quite
small (at  most  $6\%$, at highest densities); the above correction
procedure has been shown to accurately match the multiplicity function of
groups in mock catalogues \citep{ber06}; and the velocity distribution
(relative to group centre) of the original and fixed samples are
consistent \citep{lab10}. Hence, the local galaxy density estimates take
in account the fiber collision issue.

\subsection{Absolute Magnitudes and Colors}

For the current work we consider the stacked properties of cluster and field 
galaxies (in this case, regarding local density only). We do that considering the 
radial offset (in units of $R_{200}$), absolute magnitudes, colors and
local densities of all member galaxies coming from the 152 clusters (or the field). 
Our sample consists of 5,106 bright member galaxies with $M_r \le M^*+1$ (at 
$z \le 0.100$), 1,309 faint members with $M^*+1 < M_r \le M^*+3$ (at $z \le 0.045$), 
2,936 bright field galaxies and 1,740 faint field
galaxies. The bright and faint regimes are the same for cluster and field
objects.

To compute the absolute magnitudes of each galaxy (in $ugri$ bands) we consider the 
following formula: $M_x = m_x - DM - kcorr - Qz$ ($x$ is one of the four SDSS bands
we considered), where DM is the distance modulus (considering the redshift of each 
galaxy), $kcorr$ is the k$-$correction and $Qz$ ($Q = -1.4$, \citealt {yee99}) 
is a mild evolutionary 
correction applied to the magnitudes (for each galaxy redshift). The k$-$corrections 
are obtained directly from the SDSS database, for every object in each band. 
Rest-frame colors are also derived for all objects. All the magnitudes
we retrieved from the SDSS are de-reddened model magnitudes (see paper I).

\subsection{The stellar population properties}

A large number of galaxy properties were derived by different research groups
for the SDSS data set. These parameters (such as stellar mass and star
formation rate) are derived from spectral energy distribution (SED) fitting of
stellar population synthesis models, considering the galaxy spectra or the
broad band galaxy photometry. In the current work we consider parameters
derived from the application of the STARLIGHT code \citep{cid05}, the
``Galspec'' analysis provided by the MPA-JHU group (from the Max Planck
Institute for Astrophysics and the Johns Hopkins University;
\citealt{bri04}), and the MAGPHYS code \citep{dac08, dac13}
applied to the SDSS plus WISE photometry \citep{cha15}. A brief
description of the parameters we selected from these different methods is
given below.

\subsubsection{STARLIGHT}

This code fits an observed spectrum with a linear combination of a number of
template spectra with known properties (see paper III for details). We applied
the code to the galaxy spectra of the galaxies in our sample. Among other
parameters derived by STARLIGHT, we consider for this work the Age (and its
dispersion) of the stellar population and the metallicity ($Z$). Both
parameters (age and $Z$) represent the mass-weighted mean values.
We have also derived a spectral classification, on which each object is
called a Seyfert, LINER or Star-Forming (SF) according to their position in
the Baldwin, Phillips \& Terlevich (BPT) diagram. Galaxies
with no significant emission lines are called
passive. When applying STARLIGHT we require at least 30\% of good spectral
coverage (no bad quality flags on). Hence, we were not able to derive the
above parameters for all galaxies in our sample. We end up with
5,060 bright member galaxies with $M_r \le M^*+1$ (at 
$z \le 0.100$), 1,303 faint members with $M^*+1 < M_r \le M^*+3$
(at $z \le 0.045$), 2,912 bright field galaxies
and 1,714 faint field galaxies. That represents $\sim$ 1$\%$ loss.

\subsubsection{MPA-JHU}

The ``galSpec'' galaxy properties from MPA-JHU were derived for the DR8
galaxy spectra (nearly all of which were in DR7). The galaxy parameters we
selected for this work are the BPT classification, the stellar mass and
the star formation rate, and some index and line measurements,
such as D$_n$(4000) and H${\delta}$.
From the BPT diagram they divide galaxies into
the ``Star Forming'', ``Composite'', ``AGN'', ``Low S/N Star Forming'',
``Low S/N AGN'',
and ``Unclassifiable'' categories. We consider the total stellar mass and
star formation rate values (SFRs). The total stellar masses are based on model
magnitudes. SFRs are computed within the galaxy fiber aperture using
the nebular emission lines as described in \citet{bri04}. Outside
of the fiber the estimates use the galaxy photometry, as in
\citet{sal07}. AGN and galaxies with weak emission lines,
have SFRs estimates from the photometry. Due to the minimum criteria required
by the MPA-JHU group ({\it {e.g.}}, redshift, S/N) not all galaxies in our
sample are matched to the MPA-JHU results. Besides that we also
require that the galaxies had the STARLIGHT parameters available. Hence,
our sample with MPA-JHU values (common to the STARLIGHT) have 4,953 bright
and 1,297 faint member galaxies, and also 2,864 bright and 1,672 faint
field galaxies. That is 97$\%$ of the original sample.

\subsubsection{MAGPHYS}

\citet{cha15} combined SDSS and WISE photometry for the full SDSS
spectroscopic galaxy sample, creating SEDs that cover
$\lambda = 0.4-22$ $\mu m$.
They used MAGPHYS to model simultaneously and consistently both the
attenuated stellar SED and the dust emission at 12 $\mu m$ and 22 $\mu m$.
We selected from their data the star formation rate, the stellar mass, the
dust attenuation parameters $\mu$ and $\tau$, as well as the dust,
12 $\mu m$ and 22 $\mu m$ luminosities. We were able to retrieve data from
the catalog of \citet{cha15} for 4,900 bright and 1,244
faint member galaxies, and also 2,847 bright and 1,671 faint field galaxies.
That is $~ $ 97$\%$ of the original sample.

\subsection{Transitional galaxy populations: Blue bulges and Red discs}

Galaxies in the local Universe can generally be split in two types depending
on their star formation activity and structure. Early-type galaxies are
redder, with little star formation and high concentration. On the contrary
late-type galaxies are bluer, show active star formation and are less
concentrated. In paper IV we separate galaxies according to three
parameters, two related to their star-formation properties ($u-r$ color and
the spectral classification $e_{class}$), and one to the structure (the
concentration index $C = R_{90}/R_{50}$). After \citet{str01} we called blue/red
galaxies those with ($u-r$) values smaller/greater than 2.2. We considered
objects with $C < 2.6$ as disc-dominated, and the rest as bulge-dominated
galaxies ($C \ge 2.6$). Passive galaxies are those with $e_{class}$ smaller
than $-$0.05, while the star-forming objects have larger values for this
parameter. In that work we characterize the variation of different populations
from the extreme field, through the outskirts of groups and clusters, up to 
the most dense regions of the Universe, the core of these structures. Our
results indicate that pre-processing in groups leads to a strong evolution 
in galaxy properties, before they are accreted by large clusters. In agreement
to  \citet{val11}, we find that local density is the main driver for galaxy
evolution and not the parent halo mass. The evolution is such that star 
formation is quenched in the group scale, but morphological transformation
is a separate process, occurring in larger galaxy systems.

In the current work we are interested on investigating the properties of
transitional galaxies, and their environmental dependence. Those are likely
objects caught in their way from the ``blue cloud'' to the ``red sequence'',
or in a few cases passive galaxies in a process of ``rejuvenation''.
We select these galaxies simply considering two of the parameters
used in paper IV ($u-r$ and $C$). We define two types, which we call
``red discs'' (objects with $u-r \ge 2.2$ and $C < 2.6$),
and ``blue bulges'' (with $u-r < 2.2$ and $C \ge 2.6$).
Additionally we call the ``regular'' {\it early} and {\it late-types} as
``red bulges'' ($u-r \ge 2.2$ and $C \ge 2.6$), and ``blue discs''
(with $u-r < 2.2$ and $C < 2.6$), respectively. We compare the properties of
these four classes below, as well as their variation from the field
to clusters.

\begin{figure*}
%\begin{figure}
\begin{center}
\leavevmode
\includegraphics[width=7.0in]{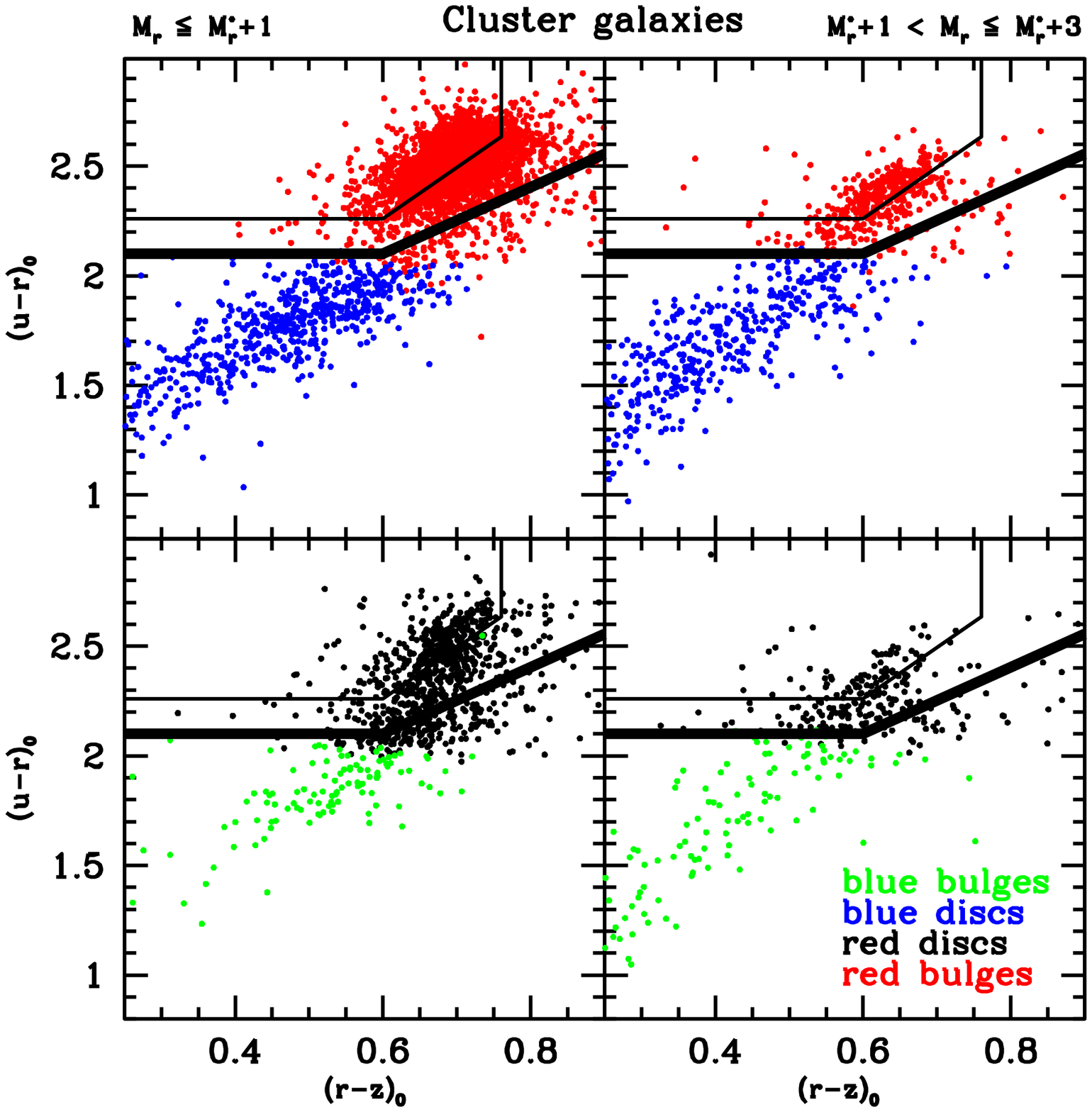}
\end{center}
\caption{Color-color diagram showing bright galaxies ($M_r \le M^*+1$)
  in the left panels and faint objects ($M^*+1 < M_r \le M^*+3$) in the
  right panels. Top panels show red bulges (red points) and blue discs (blue
  points). Bottom panels display red discs (black) and blue bulges (green).
  The thin lines show the limits adopted by \citet{hol12}, while the
  thick lines shows the limits of \citet{cha15}. Galaxies above (below)
  the lines are passive (star-forming) according to these different
  authors. Only cluster members are displayed, but the results are
  qualitatively the same for field galaxies. We can see on all panels that
  red (blue) galaxies are generally above (below) the division line
  from \citet{cha15}.}
\label{fig:gals_sep1}
\end{figure*}

As we consider two simple parameters (an observed color and a concentration
index) to separate galaxy populations there might be the question if the
samples are largely affected by other populations. For instance, some of
the ``red discs'' could be active star-forming galaxies that look red due
to large amounts of dust. To verify that is not the case we inspect the
color-color diagram, ($u-r$)$_0$ {\it vs} ($r-z$)$_0$, that is considered
an effective way for separating passive from star-forming objects
\citep{wuy07, hol12, cha15}.
Fig.~\ref{fig:gals_sep1} exhibits this diagram for the cluster galaxies
considered in this work. Bright galaxies ($M_r \le M^*+1$) are in the left
panels, while faint objects ($M^*+1 < M_r \le M^*+3$) are in the right panels. 
On top we display red bulges (red points) and blue discs (blue points), while
in the bottom panels we have red discs (black) and blue bulges (green).
The thin lines represent the limits adopted by \citet{hol12}, and the
thick lines shows the limits of \citet{cha15}. Galaxies above (below)
the lines are passive (star-forming) according to these different authors.
We can see on all panels that galaxies we considered as quiescent/star-forming
are generally above/below the division lines. In particular, the separation
given by \citet{cha15} seems more effective. We also notice the ``red
discs'' show only a small fraction of objects consistent to star-forming.

%\end{figure}

\begin{figure*}
%\begin{figure}
\begin{center}
\leavevmode
\includegraphics[width=7.0in]{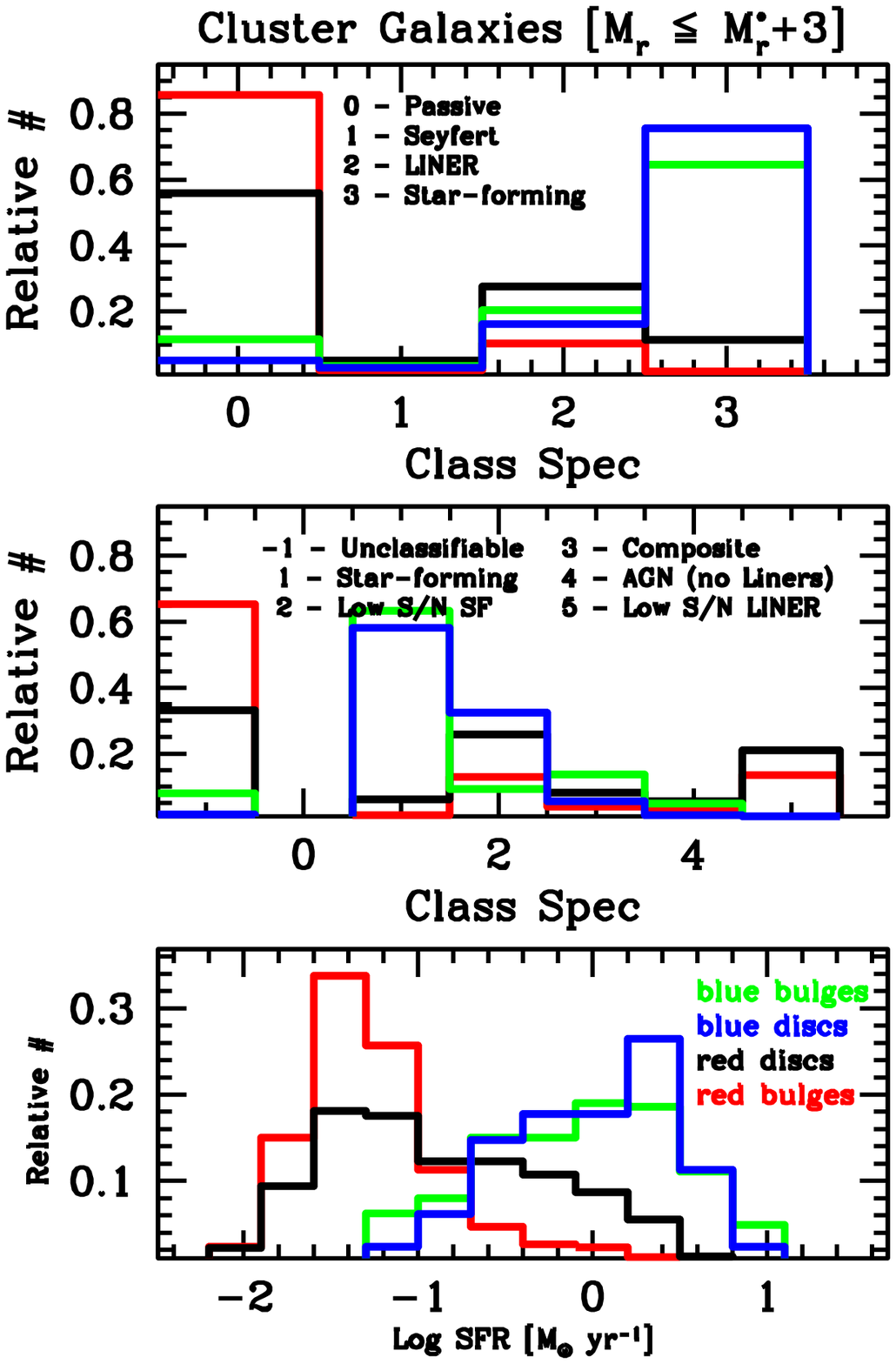}
\end{center}
\caption{Top: spectral classification from STARLIGHT. We call a galaxy
  passive if there are no significant emission lines. An object is
  classified as Seyfert, LINER or Star-Forming (SF) according to their
  position in the BPT diagram; Middle: Similar to the top panel, but
  considering the BPT diagram of the MPA group; Bottom: Star Formation Rate
  (SFR) as estimated by the MPA group. Green lines are for ``blue bulges'',
  blue lines for ``blue discs'', black lines for ``red discs'', and red lines
  for ``red bulges''. All cluster objects with $M_r \le M^*+3$ (bright
  and faint) are considered. This figure shows that most ``red discs'' are
  passive, although having a non-negligible fraction of galaxies with residual
  SF compared to the ``red bulges''. Most ``blue bulges'' are classified as SF
  objects and show similar SFR to the ``blue discs''.}
\label{fig:bptclass_sfr}
\end{figure*}

In Fig.~\ref{fig:bptclass_sfr} we confirm these conclusions through the
inspection of the spectral classification from the STARLIGHT code (top)
and the MPA-JHU group (middle panel). In the top panel a galaxy is called
passive if there are no significant emission lines. From the BPT diagram an
object is classified as Seyfert, LINER or Star-Forming (SF). In the middle
panel a more detailed classification is given, with the ``SF'', ``Low S/N SF'',
``Composite'', ``AGN'', ``Low S/N LINER'' and ``Unclassifiable'' possibilities.
Green lines are for ``blue bulges'', blue lines for ``blue discs'',
black lines for ``red discs'', and red lines for ``red bulges''. These
results consider all cluster galaxies with $M_r \le M^*+3$ (bright and faint).
As we can see from the top panel $\sim 55 \%$ of ``red discs'' are passive,
with only $\sim 10 \%$ being considered SF. The rest is most composed of
LINERs. Similar conclusions are reached from the middle panel, with
$< 10 \%$ of the ``red discs'' being considered SF, although this more refined
classification results in $\sim 25 \%$ of these galaxies being called
``Low S/N SF''.

In the bottom panel of Fig.~\ref{fig:bptclass_sfr} we display the
Star Formation Rate (SFR) distribution estimated by the MPA group for these
four populations. From this plot we can see the ``red discs'' are clearly
distinct from the regular ``blue discs''. However, the SFR distribution of the
``red discs'' is not identical to the the ``red bulges'' neither. The former
shown a non-negligible fraction of galaxies with residual SF in comparison to
the latter. Note that the blue galaxies (bulges or discs) have more similar
SFR distributions.

%\end{figure}

In order to have a rough idea of the morphological
classification of the galaxies (especially the transition
populations) we decided to perform a visual inspection of galaxies
in our sample. That was done by BV for the ``red disc'' and ``blue bulge''
bright populations. A total of 1332 red discs were
inspected, 836 within clusters and 496 in the field. For the blue bulge
population we have 111 member galaxies and 157 field objects. The red discs
were classified as ``face-on red spiral'' ({\it frs}),
``face-on red/blue spiral'' ({\it fbs}, with significant bluer colors
in the arms), ``edge-on disc'' ({\it edg}), and ``spheroidal'' ({\it sph}).
The blue bulges were classified as ``spheroidal'' ({\it sph}),
``edge-on disc'' ({\it edg}),  ``face-on disc'' ({\it fd}), and
``double core'' ({\it dc}, for galaxies with two cores
or strong signs of interaction). From these visual morphological
classifications we verified that more than $50 \%$ of
the ``red discs'' are classified as
{\it frs} (``face-on red spiral'') and less than $20 \%$ are called
{\it edg} (``edge-on disc''), reinforcing that misclassification due to
edge-on spirals with high star formation is not an issue. The rest is
classified as {\it fbs} ($\sim$ $10 \%$) or {\it sph} ($\sim$ $20 \%$).
For the ``blue bulges'' more than $50 \%$ are classified as {\it sph}
(``spheroidal''). The remaining objects are classified as {\it fd}
($\sim$ $35 \%$), {\it edg} ($\sim$ $15 \%$), or {\it dc} ($\sim$ $5 \%$).

Note this classification process is very uncertain as even for the bright
galaxies a high level of agreement between different classifiers is
usually not achieved. That is due to the subjective nature of such visual
inspection and low contrast of many stamps. Our main goal with such exercise
was to confirm the expectations from Figs.~\ref{fig:gals_sep1} and
\ref{fig:bptclass_sfr}, roughly verifying the ``red discs'' are not
dominated by ``edge-on discs'' and the ``blue bulges'' are mostly classified
as ``spheroidal'' systems. It is also important to mention the reason we
did not extended such visual inspection for the fainter objects is the fact
that even for some bright systems the classification was already very
difficult. As a matter of fact, our first step was trying to use the
classification from the ``Galaxy Zoo'' project \citep{lin08}, both for
bright and faint galaxies of our sample.
The ``Galaxy Zoo'' project considers an object as ``spiral'' if the debiased
spiral fraction is larger than 0.8. If the debiased elliptical fraction is
larger than 0.8 then the galaxy is classified as ``elliptical''. If none of
those cases are true the galaxy is called ``uncertain''. Considering those
three possibilities we had to discard $\sim$ $60 \%$ of our sample.
Unfortunately, all these objects are called ``uncertain''. Due to that we did
not use the classification from ``Galaxy Zoo''. It is important to stress that
in the current work we are mainly interested on comparing the physical
properties and environment of galaxy populations divided by color and
concentration, and not on performing a detailed morphological
classification. Hence, we consider the morphological inspection we performed
is satisfactory for our goals.

%CC\begin{figure}
%CC	\includegraphics[width=\columnwidth]{example}
%CC    \caption{This is an example figure. Captions appear below each figure.
%CC	Give enough detail for the reader to understand what they're looking at,
%CC	but leave detailed discussion to the main body of the text.}
%CC    \label{fig:example_figure}
%CC\end{figure}

\subsubsection{Possible systematic effects as a function of axial ratio}

Although the ``red disc'' population seem not to be contaminated by
a large number of highly elongated dusty spirals we decided to perform one
more check based on the galaxy's axial ratio (taking $a/b$ as a proxy for
inclination). As done by \citet{mas10} and \citet{toj13} we consider an axial
ratio limit of log ($a/b$) $ = $ 0.2 to separate the low and high inclination
objects. We compared a few physical properties (such as the sSFR) of disc
galaxies with log ($a/b$) $ < $ 0.2 and log ($a/b$) $ > $ 0.2, doing so for the
``red disc'' and ``blue disc'' populations. We noticed the sSFR of low
and high inclination ``red discs'' are indeed different, but the
``blue discs'' are not affected by a division in axial ratio. However,
the two populations (red and blue discs) are still completely distinct.
As we are most interested on investigating the differences between red and
blue discs, we decided not to impose an axial ratio cut to our sample.
We found this cut on axial ratio does not affect any of our results and
conclusions through the paper. For instance, the comparison of ``blue discs''
and ``red discs'' in Fig.~\ref{fig:ZagesSFR_mass} (showing the relations
between stellar metallicity, the age of the stellar population, and
the sSFR {\it vs} stellar mass) is not affected by
a restriction to galaxies with log ($a/b$) $ < $ 0.2. Hence, as the
environmental variation of the populations or their differences are
not sensitive to an axial ratio cut we did not impose such restriction
to our sample.

\section{The Environmental Variation of Transitional Galaxy Populations}

In this section we investigate the variation of the two transitional galaxy
populations (``red discs'' and ``blue bulges'') with the environment,
characterized by the local galaxy density, and the normalized radial distance
to the parent cluster. We have also investigated the environmental influence
in different ranges of galaxy stellar mass. We show the variation of
these populations at fixed morphology. For instance, we verify how the
fraction of ``red discs'' relative to all discs depends on environment
(as well as the fraction of ``blue bulges'' relative to all bulges).

\subsection{Variation with Local Galaxy Density}

\begin{figure}
\begin{center}
\leavevmode
\includegraphics[width=3.5in]{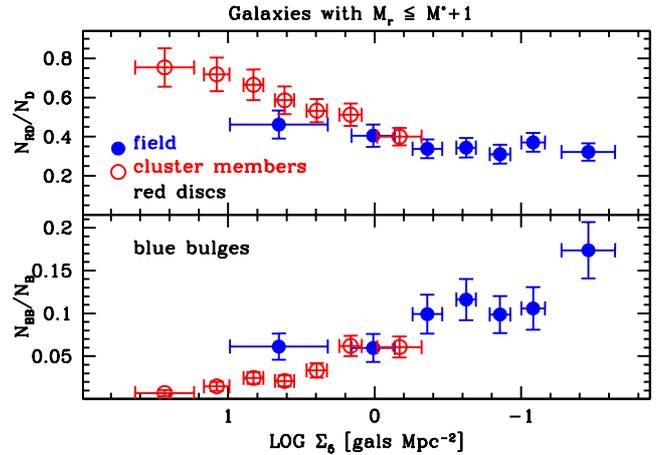}
\end{center}
\caption{Fraction of red discs (top) relative to the full disc population,
  and the fraction of blue bulges (bottom) relative to all bulges. Both
  fractions are displayed as a function of local galaxy density ($\Sigma_5$).
  These results consider bright galaxies ($M_r \le M^*+1$) only. At fixed
  morphology, from low to high density the proportion of ``blue bulges''
  decreases. For the ``red discs'' the relation is nearly flat in the field,
  increasing steeply in the cluster domain.}
\label{fig:denbins2}
%\end{figure*}
\end{figure}

\begin{figure}
\begin{center}
\leavevmode
\includegraphics[width=3.5in]{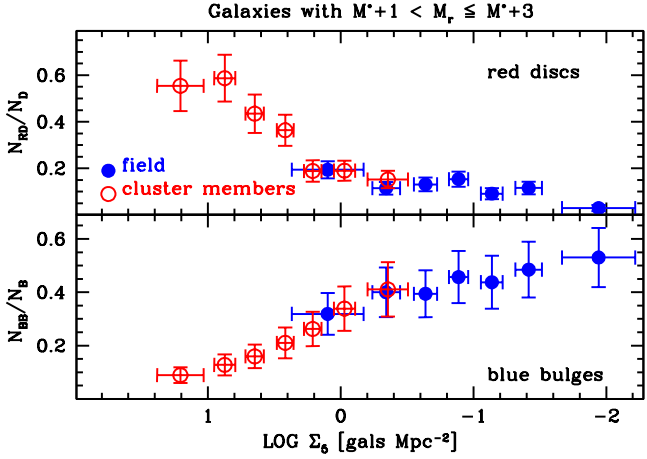}
\end{center}
\caption{Analogous to Fig.~\ref{fig:denbins2}, but showing the results for
  faint galaxies ($M^*+1 < M_r \le M^*+3$) only. These two figures represent
  the colour-density relation, showing that at fixed morphology (discs) the
  proportion of red galaxies increases with density.}
\label{fig:denbinsb2}
%\end{figure*}
\end{figure}

In Figs.~\ref{fig:denbins2} and ~\ref{fig:denbinsb2} we show the variation in
the number of ``red discs'' over all discs (top), and ``blue bulges'' relative
to all bulges (bottom), as a function of local galaxy density ($\Sigma_5$).
Fig.~\ref{fig:denbins2} is for bright galaxies ($M_r \le M^*+1$) and
Fig.~\ref{fig:denbinsb2} for faint objects ($M^*+1 < M_r \le M^*+3$).
Red open circles show cluster members, while blue filled circles indicate
field galaxies. From the low to the high density
regime the number of ``blue bulges'' (relative to all bulges) decreases,
especially once within clusters. That is true for bright and faint objects.
For the ``red discs'' we detect a nearly flat relation for
field galaxies and a steep increase in the relative number to the disc
population, both for bright and faint objects. The number of ``blue bulges''
over bulges varies from $\sim 1 \%$ to $\sim 18 \%$ (for bright galaxies),
and from $\sim 10 \%$ to $\sim 50 \%$ (for faint objects). The number of
``red discs'' over discs changes from $\sim 35 \%$ to $\sim 80 \%$
(for the bright regime), and from $\sim 3 \%$ to $\sim 60 \%$ (for faint
galaxies).

A result we should highlight is the fact disc galaxies become redder as
density increases, but this effect is mainly seen in the cluster environment.
In particular, as we are going to see below (Figs.~\ref{fig:radbins2} and
~\ref{fig:radbinsb2}), this transformation occurs mainly within the cluster
virial radius. That is another way to represent the colour-density
(or colour-radius) relation. At fixed morphology (``discs'') we detect a steep
variation in the number of ``red discs'' over all ``discs'' with density
(once within the clusters). Hence, those ``blue discs'' change their
colour - becoming red - while still keeping the same morphology.

%\begin{figure*}

%\begin{figure*}

\subsection{Dependence on the Cluster Radial Distance} 

\begin{figure}
\begin{center}
\leavevmode
\includegraphics[width=3.5in]{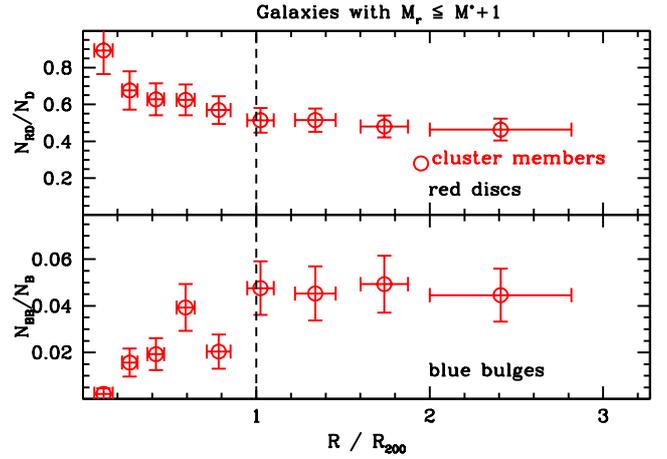}
\end{center}
\caption{Fraction of red discs (top) and blue bulges (bottom) as a function
  of the normalized distance to the cluster center ($R/R_{200}$). Only cluster
  members are shown. These results consider bright galaxies ($M_r \le M^*+1$)
  and the fractions are relative to the full disc population (top), and to
  all bulges (bottom). The vertical dashed line indicates the $R_{200}$
  radius. From the outskirts to the inner parts of the clusters a large
  variation is seen only once within the virial radius (approximated by
  $R_{200}$).}
\label{fig:radbins2}
%\end{figure*}
\end{figure}

\begin{figure}
\begin{center}
\leavevmode
\includegraphics[width=3.5in]{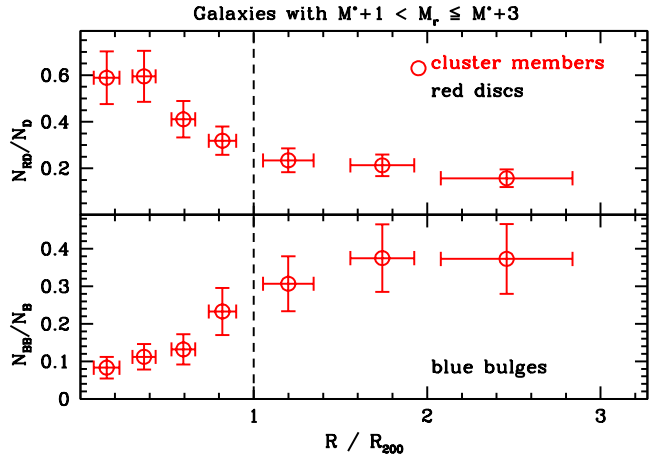}
\end{center}
\caption{Same as previous figure, but showing results for faint galaxies 
  ($M^*+1 < M_r \le M^*+3$). From the outskirts to the center of clusters
  the relative numbers of the two populations (``blue bulges'' and
  ``red discs'') change mostly once within $R_{200}$. These results
  reinforce the importance of the cluster environment to transform
  galaxy properties.}
\label{fig:radbinsb2}
%\end{figure*}
\end{figure}

In Figs.~\ref{fig:radbins2} and ~\ref{fig:radbinsb2} we show the variation in
the number of ``red discs'' over all discs (top), and ``blue bulges'' relative
to all bulges (bottom), as a function of the normalized distance to the
parent cluster ($R/R_{200}$). The results are only shown to cluster members.
Fig.~\ref{fig:radbins2} is for bright galaxies ($M_r \le M^*+1$) and
Fig.~\ref{fig:radbinsb2} for faint objects ($M^*+1 < M_r \le M^*+3$). The
symbols are the same as in Figs.~\ref{fig:denbins2} and ~\ref{fig:denbinsb2},
for cluster galaxies. From the outskirts to the inner parts of the clusters
we see the two populations (``blue bulges'' and ``red discs'') are nearly
constant until within $R_{200}$, and then the relative numbers decrease
(increase) for the ``blue bulges'' (``red discs'') all the way to the core.
That is true for bright and faint galaxies. These results reinforce the
relevance of the cluster environment on transforming galaxy properties,
especially those related to the star formation activity.

%\begin{figure*}

%\begin{figure*}

\subsection{Dependence on the Galaxy Stellar Mass} 

\begin{figure}
\begin{center}
\leavevmode
\includegraphics[width=3.5in]{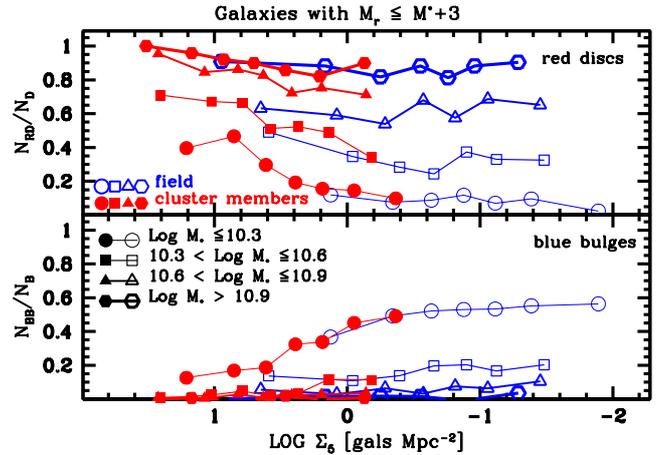}
\end{center}
\caption{Same as in Fig.~\ref{fig:denbins2}, but splitting the sample
  according to the galaxy stellar mass. We consider four intervals:
  $Log M_* \le 10.3$, $10.3 < Log M_* \le 10.6$,
  $10.6 < Log M_* \le 10.9$, $Log M_* > 10.9$. From low to high stellar
  mass the line used to connect the points is progressively thicker.
  As before, cluster members results are in red and field results in blue.
  All galaxies with $M_r \le M^*+3$ (bright and faint) are considered.
  We can see that star formation is halted first in higher mass galaxies.
  For the ``red discs'' the largest variation is seen for the lower mass
  objects ($Log M_* \le 10.6$), while for the ``blue bulges'' a significant
  change is detected only for the lowest mass galaxies ($Log M_* \le 10.3$).}
\label{fig:den_stmassbins2}
%\end{figure*}
\end{figure}

In Fig.~\ref{fig:den_stmassbins2} we show again the variation in
the number of ``red discs'' over all discs (top), and ``blue bulges'' relative
to all bulges (bottom), as a function of the local galaxy density
($\Sigma_5$). This figure is analogous to Fig.~\ref{fig:denbins2}, but now
we show the results for bright and faint galaxies at the same time
($M_r \le M^*+3$), as we consider four different stellar mass ranges:
$Log M_* \le 10.3$ (circles),
$10.3 < Log M_* \le 10.6$ (squares),
$10.6 < Log M_* \le 10.9$ (triangles), and
$Log M_* > 10.9$ (hexagons). From low to high stellar mass the line used
in Fig.~\ref{fig:den_stmassbins2} is progressively thicker.
Cluster members are show in red (filled symbols), while field galaxies
are in blue (open symbols).

As expected, the largest fractions of red galaxies are seen for the higher
mass galaxies (the last two mass bins, $Log M_* > 10.6$). However, the
dependence on $\Sigma_5$ is smaller for these massive galaxies when compared
to the first two mass bins ($Log M_* \le 10.6$), for which we see a
steep increase in the number of ``red discs'' over discs, especially within
clusters. On what regards the ``blue bulges'' the variation with $\Sigma_5$
is significant only for the lowest mass galaxies ($Log M_* \le 10.3$).
These results reinforce the idea that star formation is halted first in
higher mass galaxies. The fact the massive field ``red discs'' have a
negligible variation with local density, but their cluster counterpart still
depend on $\Sigma_5$ (increasing from $\sim 80 \%$ to $\sim 100 \%$) stress
the importance of the cluster environment to transform these objects.

%\begin{figure*}

\section{Environmental and Physical Properties of the Galaxy Populations}

In this section we compare some physical properties (such as age and
metallicity) of the four galaxy populations defined for this work
(``blue bulges'', ``blue discs'', ``red discs'' and ``red bulges''), as
well as some properties related to the environment ($R/R_{200}$,
crossing time, and $\Sigma_5$). This comparison is made with the help of
the cumulative distributions of these properties. Our goal is to verify
if the galaxy populations can really be considered as different types of
objects or, in other words, systems at different evolutionary stages.
The crossing time (t$_{cross}$) is defined
as the distance of the galaxy to the group center divided by the group's
line-of-sight velocity dispersion. For a large sample of galaxies
t$_{cross}$ provides a measure of how long galaxies have been affected
by the group environment \citep{lac13}.

\begin{figure*}
%\begin{figure}
\begin{center}
\leavevmode
\includegraphics[width=7.0in]{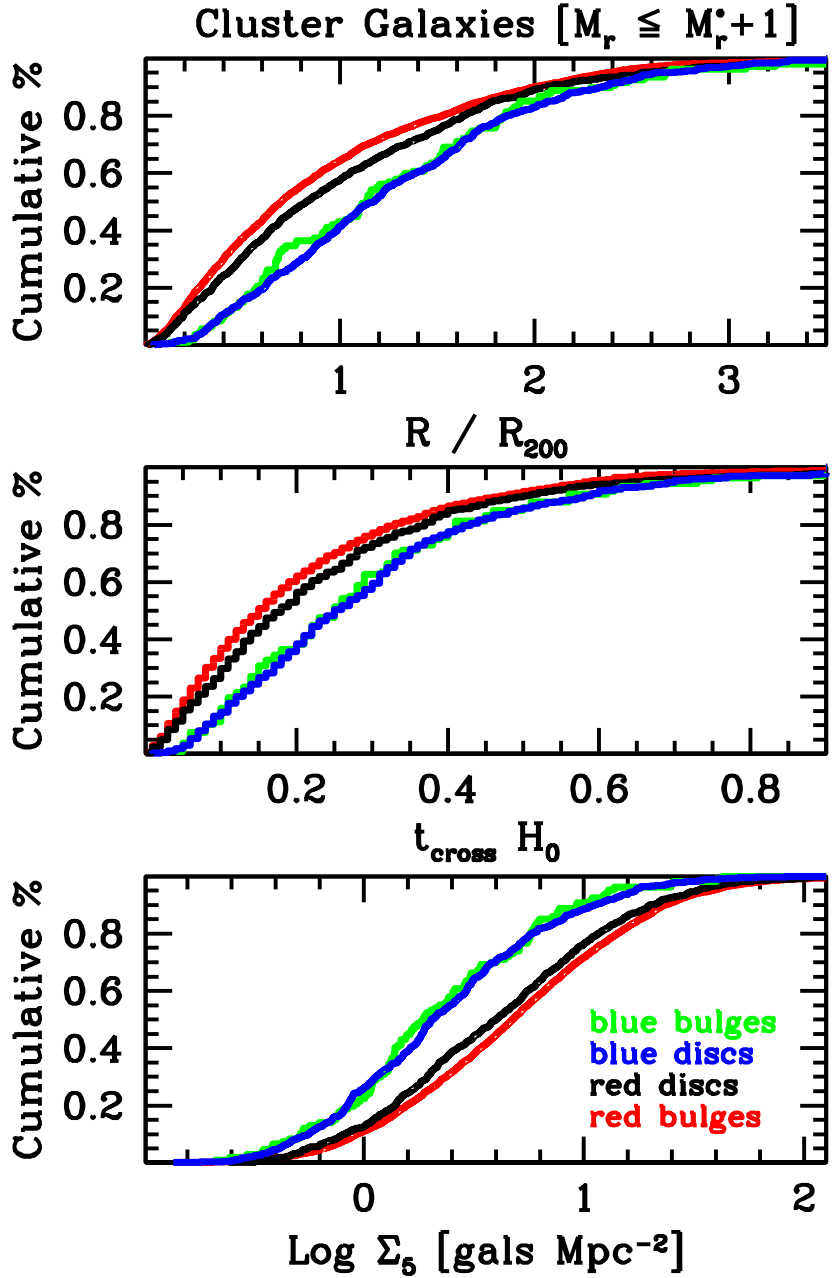}
\end{center}
\caption{Cumulative distributions for different galaxy populations:
  ``blue bulges'' in green, ``blue discs'' in blue, ``red discs'' in black,
  and ``red bulges'' in red. The top panel shows the normalized distance
  to the cluster center ($R/R_{200}$), while the middle panel displays
  the crossing time, and the local galaxy density is exhibited in the bottom
  panel. Only bright cluster galaxies ($M_r \le M^*+1$) are considered. At
  fixed morphology red and blue discs are found in different environments
  (the same is true for the bulges, red and blue). On the contrary, at
  fixed color the galaxy environment is similar, for bulges or discs.}
\label{fig:rad_tc_sig5}
\end{figure*}

Fig.~\ref{fig:rad_tc_sig5} shows the cumulative distributions of these four
populations, with ``blue bulges'' in green, ``blue discs'' in blue,
``red discs'' in black, and ``red bulges'' in red (as in
Fig.~\ref{fig:bptclass_sfr}). The normalized distance to the cluster center
($R/R_{200}$) is exhibited on top, the crossing time (shown in units of
Hubble time) is in the middle panel, and $\Sigma_5$ is in the bottom panel.
Only bright cluster galaxies ($M_r \le M^*+1$) are considered.
We can see there is little difference
in the typical environment of red galaxies (discs and bulges). The same can
be said for the blue objects. However, at fixed morphology we see clear
differences between red and blue discs, as well as bulges. As it is well
known red objects are generally found at the highest densities, in the central
parts of clusters, and have shorter crossing times than blue galaxies. As
we see from Fig.~\ref{fig:rad_tc_sig5} red and blue discs are located in
very different environments, reinforcing the idea these two populations
are different. Hence, as we suggested above, the ``red disc'' population
is not simply the result of contamination of disc galaxies that look redder
due to dust emission.

\begin{figure*}
%\begin{figure}
\begin{center}
\leavevmode
\includegraphics[width=7.0in]{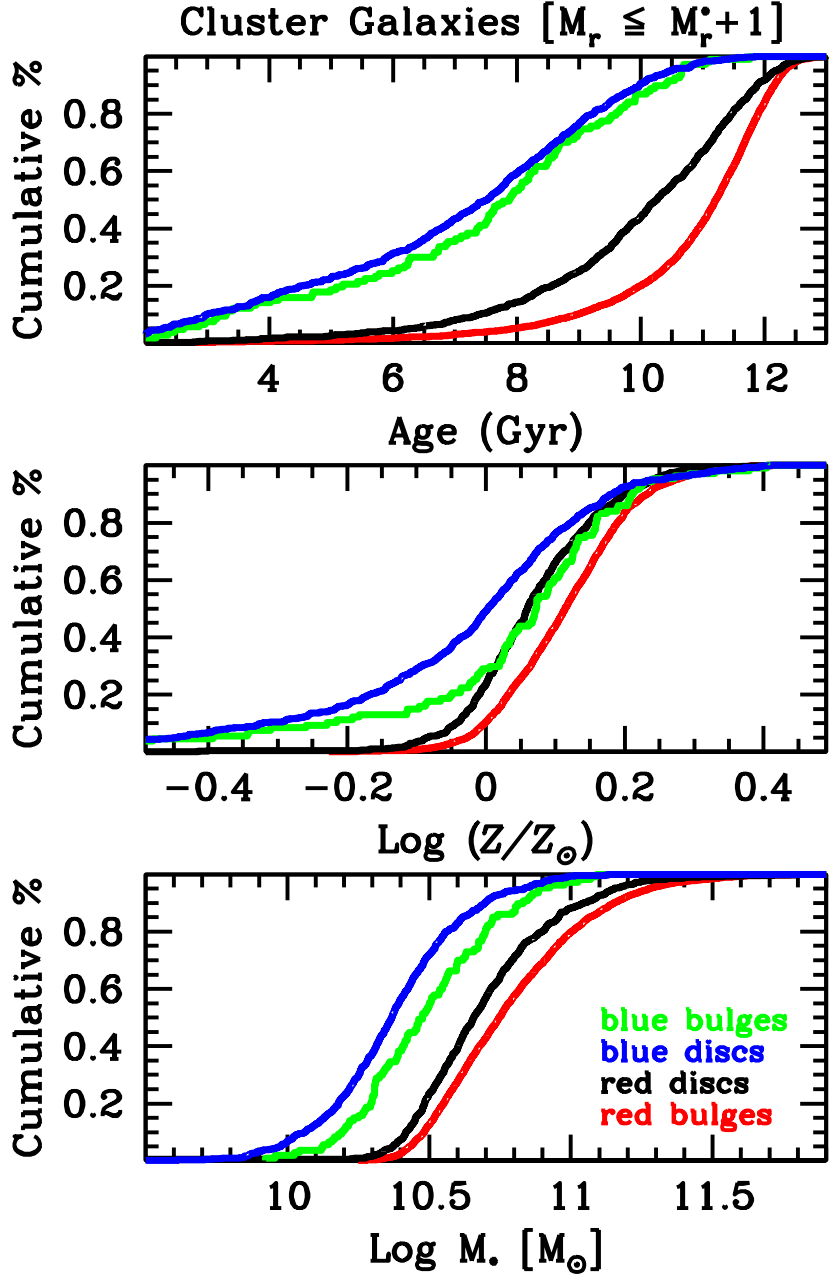}
\end{center}
\caption{Analogous to the previous figure, but showing the age of the
  stellar population (top), the metallicity (middle) and the stellar mass
  (bottom). Only bright cluster galaxies ($M_r \le M^*+1$) are considered.
  The color codes are the same as in the previous figure. At fixed morphology
  we detect a large difference in these physical properties. Hence, red and
  blue discs (or bulges) can be considered as distinct populations.}
\label{fig:age_metl_mass}
\end{figure*}

Fig.~\ref{fig:age_metl_mass} is analogous to Fig.~\ref{fig:rad_tc_sig5}, but
shows the cumulative distributions of the age of the stellar population
(top), the stellar metallicity ($Z$) in the middle panel, and the stellar
mass (bottom panel). The color codes are the same as in
Fig.~\ref{fig:bptclass_sfr} and will be kept for future
plots from now on. As above only bright cluster
galaxies are shown. We see a large
difference in these physical properties of galaxies at fixed morphology
(red and blue discs, or bulges). However, we also see a significant difference
in the properties of red galaxies (discs and bulges), especially for the two
top panels (age and metallicity). Hence, despite the fact these objects are
found at similar environments (see Fig.~\ref{fig:rad_tc_sig5}) they have
different ages and metallicities. ``Red bulges'' have an older stellar
population and higher metallicities than ``red discs''. The stellar masses of
the former are also slightly higher than the latter.

Another feature we detect in Fig.~\ref{fig:age_metl_mass} is the metallicity
difference between blue discs and bulges. Although, these two populations
are found in similar environments and have similar age distributions the
``blue bulges'' display higher $Z$ values than the ``blue discs'', and
actually get close to the results for the ``red discs''. Part of
this effect is explained by the slightly larger masses of the ``blue bulges''.
But the main reason is environmental. Fig.~\ref{fig:age_metl_mass} shows the
results only for bright cluster galaxies. For the field data we find the
metallicity distributions of blue discs and bulges to be consistent to each
other, and different than the $Z$ values of red galaxies. The same is true
if we also include the fainter galaxies. This environmental difference may
be related to different types of galaxies composing the ``blue bulge''
population in the clusters compared to the field. In $\S$5.3 we investigate
asymmetry in the ``blue bulge'' population. From the asymmetry distribution
and visual inspection we find that the ``blue bulges'' can be simple spheroids,
or bulge dominated low mass spirals, or wet mergers.
We believe that different fractions
of these types (among the ``blue bulges'') can explain the metallicity
difference between ``blue bulges'' in the field and clusters. Further
investigations are left for future work.

\begin{figure*}
%\begin{figure}
\begin{center}
\leavevmode
\includegraphics[width=7.0in]{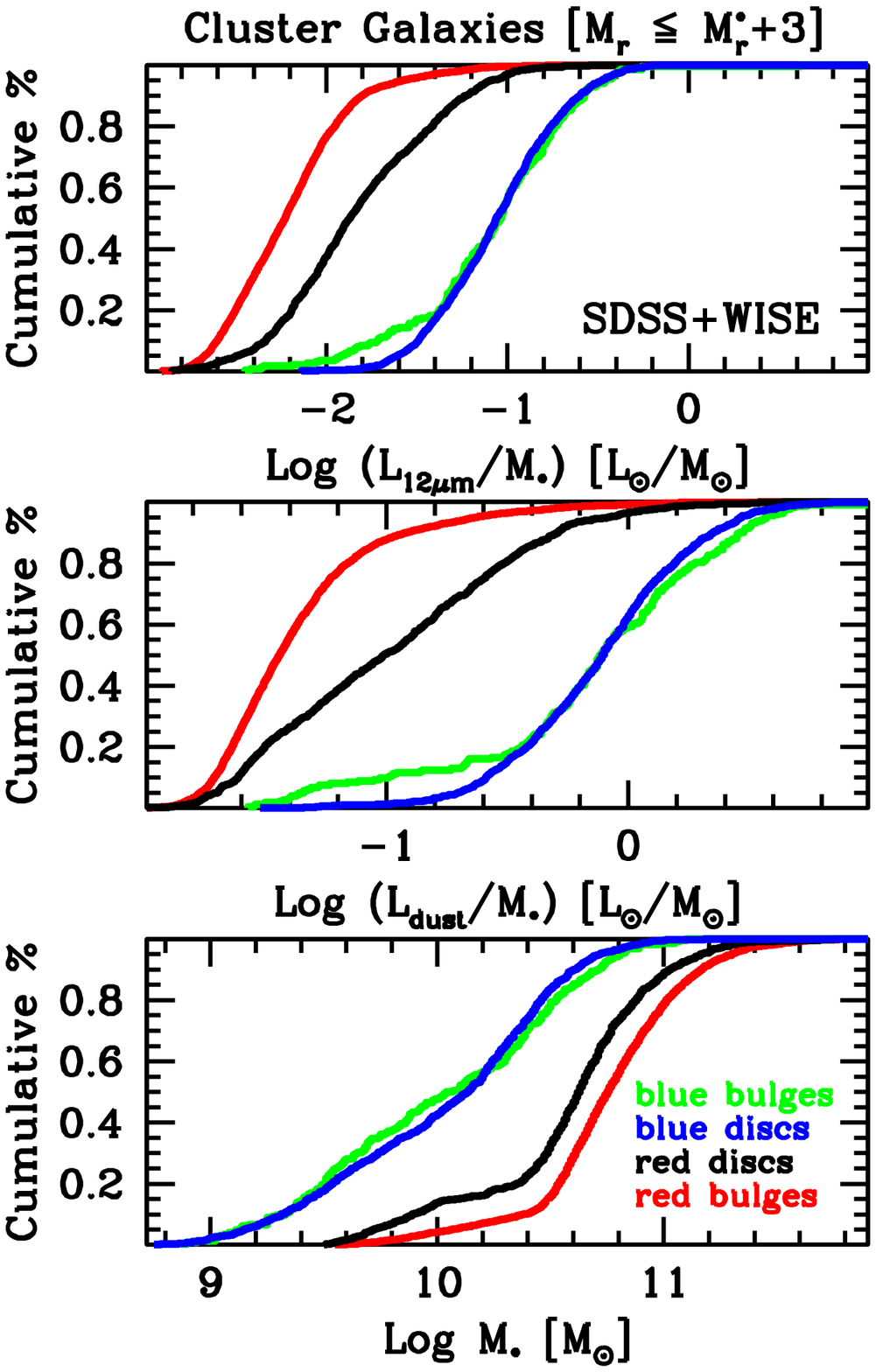}
\end{center}
\caption{Analogous to Fig.~\ref{fig:age_metl_mass}, but showing the results
  derived from the MAGPHYS code applied to SDSS+WISE data. We exhibit the
  $12 \mu m$ luminosity divided by stellar mass (top),
  the dust luminosity divided
  by mass (middle) and the stellar mass (bottom).
  All cluster galaxies ($M_r \le M^*+3$) are considered. The color
  codes are the same as in Fig.~\ref{fig:age_metl_mass}. The $12 \mu m$
  and dust luminosities point to higher star formation values for the two
  blue populations, but still indicates a residual star formation for the
  ``red discs''.}
\label{fig:sl12_dustl_mass}
\end{figure*}

The age, metallicity, stellar mass
and SFR displayed in Figs.~\ref{fig:age_metl_mass} and
\ref{fig:bptclass_sfr} were derived from STARLIGHT and the MPA-JHU group
(based only on SDSS data). Now we investigate the results obtained with
parameters derived from the MAGPHYS code applied to SDSS plus WISE data.
Fig.~\ref{fig:sl12_dustl_mass} is analogous to Fig.~\ref{fig:age_metl_mass},
but exhibiting the $12 \mu m$ luminosity
divided by stellar mass (top), the dust
luminosity divided by mass (middle) and the stellar mass (bottom). All cluster
galaxies ($M_r \le M^*+3$) are considered. From the bottom panel we can
see the stellar mass distributions are very similar
to the ones derived from the MPA-JHU group (Fig.~\ref{fig:age_metl_mass}),
but now considering the faint regime. As
noticed by \citet{cha15} adding the four WISE bands do not lead to
different stellar masses estimates. However, they verify the SFR estimates
are very different. We decided not to perform this comparison in the current
work. However, we can say that when also using WISE data the SFR values become
larger for the star-forming (blue galaxies), and smaller for the passive
objects (red galaxies). The central panel of Fig.~\ref{fig:sl12_dustl_mass}
shows that ``blue bulges'' have similar dust luminosities than ``blue discs'',
and the latter have much larger L$_{dust}$ values than ``red discs''. The
``red bulges'' have even lower dust luminosities. The top panel of this
figure reinforces what is seen in the bottom panel of
Fig.~\ref{fig:bptclass_sfr}. As the $12 \mu m$ luminosity is a good SFR
indicator \citep{cha15} we can see the most luminous galaxies in this band
are the blue ones, while the less luminous are the red galaxies (with the
``red bulges'' being the less luminous of all). Hence, the $12 \mu m$
luminosity indicates a high star formation rate for the two blue populations
(bulges and discs), while still pointing to a residual star formation for the
``red discs''.

%\end{figure}

%\end{figure}

%\end{figure}

\section{Role of Stellar Mass, Local and Global Environment}

We now want to disentangle the impact of the galaxy stellar mass and the
environment (local and global) for the physical properties of the different
galaxy populations.

\begin{figure*}
%\begin{figure}
\begin{center}
\leavevmode
\includegraphics[width=7.0in]{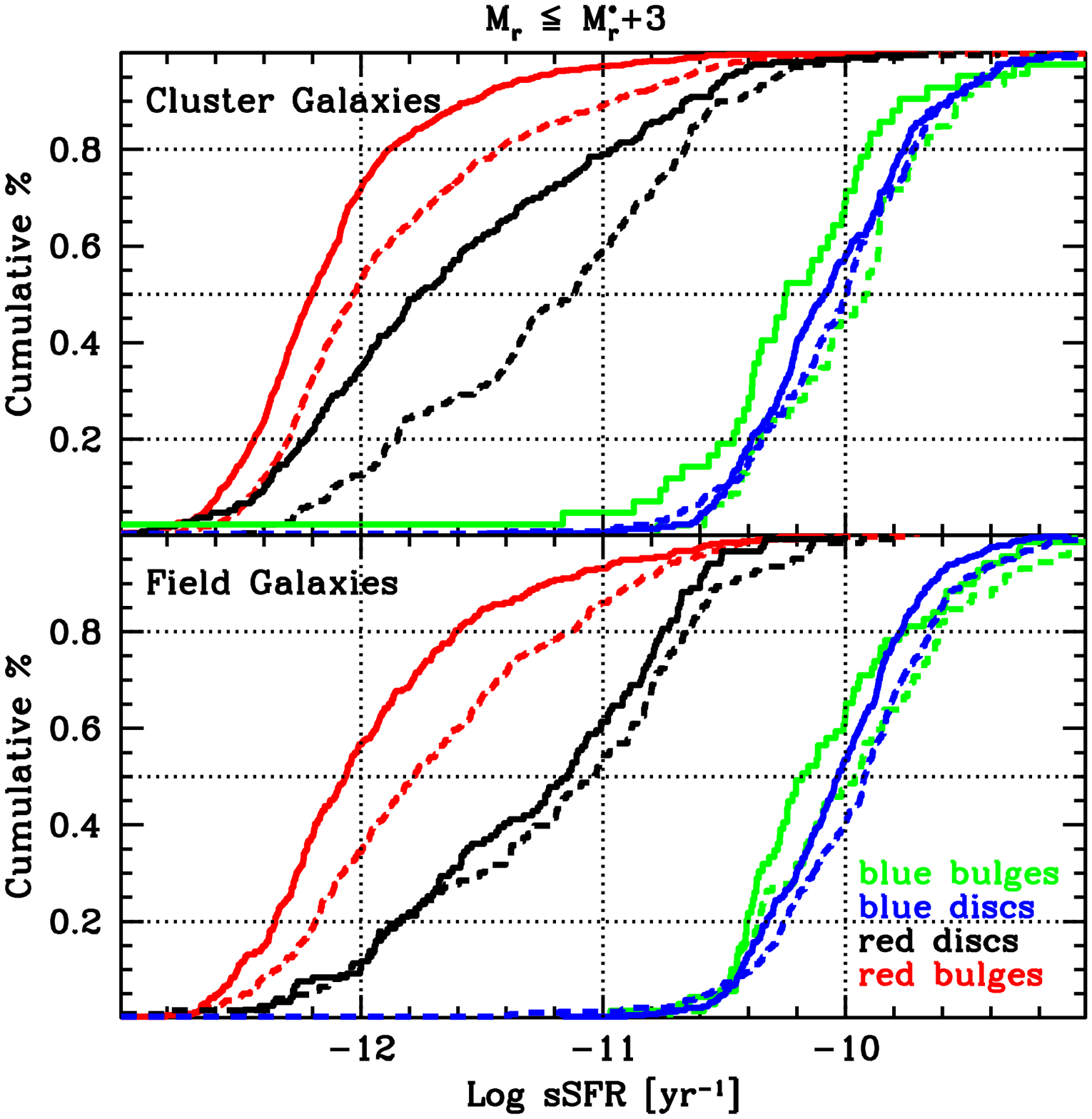}
\end{center}
\caption{Analogous to Fig.~\ref{fig:age_metl_mass},
  but showing the results only for the specific star formation rate, and in
  different local galaxy density bins. Each galaxy population (``blue bulges'',
  ``blue discs'', ``red discs'' and ``red bulges'') is divided in five ranges
  of galaxy density, with the same number of objects.
  However, to avoid confusion, we only
  exhibit the lower (dashed) and upper density (solid) cumulative
  distributions, for each population. Results for cluster galaxies are shown
  in the top panel, and for the field in the bottom panel. For this figure all
  galaxies with $M_r \le M^*+3$ (bright and faint) are considered. The color
  codes are the same as in Fig.~\ref{fig:age_metl_mass}. From this figure we
  conclude the blue and red populations do not overlap; ``red discs'' and
  ``red bulges'' are distinct objects; the blue objects display similar
  distributions; the largest variations with local density (at fixed
  morphology) are found for the red objects; and field and cluster results
  show no large differences (except for the red objects at high density).}
\label{fig:ssfr_densbins}
\end{figure*}

First we show in Fig.~\ref{fig:ssfr_densbins} the cumulative distributions
of the specific star formation rate (sSFR) in different ranges of local
galaxy density. Each galaxy population (``blue bulges'', ``blue discs'',
``red discs'' and ``red bulges'') is divided in five ranges of galaxy density,
with the same number of objects. However,
to avoid confusion, we exhibit only the
lower (dashed) and upper density (solid) cumulative distributions, for each
population. In the top panel we present the results for cluster galaxies and
in the bottom panel for the field. Doing so we can assess the importance of
the local environment (through the local galaxy density) and the global
environment (comparing field and cluster results). For this plot all galaxies
with $M_r \le M^*+3$ are considered. 

We can derive a few interesting conclusions from this figure. First the blue
and red populations do not overlap. Second, the ``red discs'' and
``red bulges'' are distinct objects both at low and high local density values.
Even the highest density results for the ``red discs'' do not overlap the
lower density curve of ``red bulges''. Third, the blue objects show similar
distributions, with the ``blue discs'' being insensitive to local
environment, and the ``blue bulges'' displaying different distributions with
local density. However, the largest variations with local density are seen
for the red objects at fixed morphology, especially for the ``red discs''
in the cluster environment. Four, the comparison of field and cluster results
show no significant variation, except for the red objects distributions
at high $\Sigma_5$ (solid curves).

%\end{figure}

\begin{figure*}
%\begin{figure}
\begin{center}
\leavevmode
\includegraphics[width=7.0in]{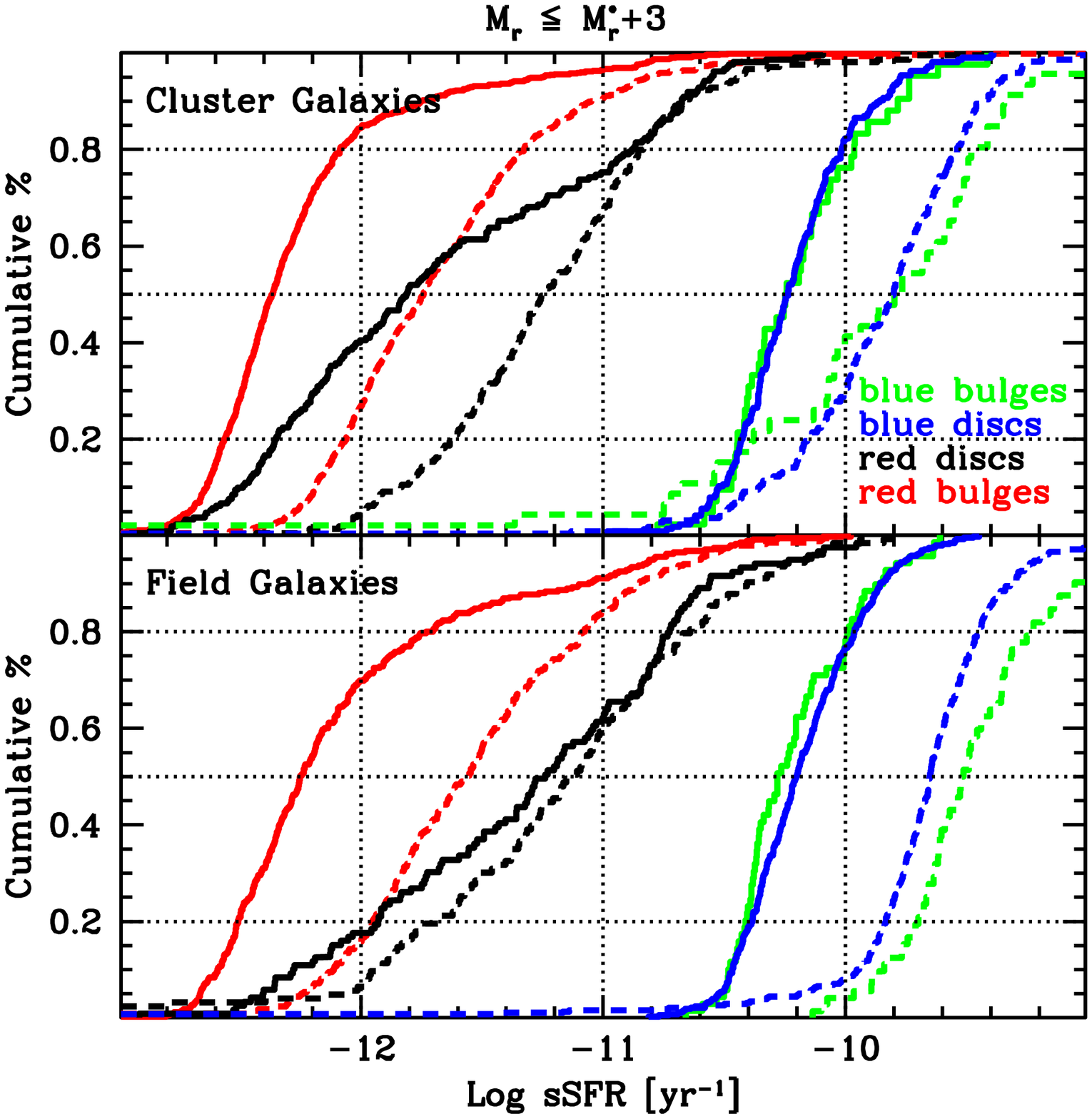}
\end{center}
\caption{Analogous to the previous figure, but showing the results for
  different stellar mass bins. Each galaxy population (``blue bulges'',
  ``blue discs'', ``red discs'' and ``red bulges'') is divided in five ranges
  of stellar mass, with the same number of objects. However, we only
  exhibit the lower (dashed) and upper mass (solid) cumulative
  distributions for each population. Results for cluster galaxies are shown
  in the top panel, and for the field in the bottom panel. For this figure all
  galaxies with $M_r \le M^*+3$ (bright and faint) are considered. The color
  codes are the same as in Fig.~\ref{fig:age_metl_mass}. All the four
  populations show a strong dependence on stellar mass (larger than what
  has been seen for local density). Hence, the most important parameter
  affecting all the four galaxy populations is the stellar mass, both in
  the field and clusters.}
\label{fig:ssfr_massbins}
\end{figure*}

As a comparison we show in Fig.~\ref{fig:ssfr_massbins} the cumulative
distributions of the specific star formation rate (sSFR), but now in
different ranges of stellar mass. As before, all galaxies with
$M_r \le M^*+3$ are considered. Each of the four galaxy populations is
divided in five ranges of stellar mass, with the same number of objects.
However, we show only the lower (dashed) and upper mass (solid)
cumulative distributions. In the top panel we show cluster results, while
in the bottom we display the field distributions. Some results are similar
to what is seen in Fig.~\ref{fig:ssfr_densbins}. For instance, the blue and
red populations show distinct distributions, both in the field and clusters.
However, now we see that all the four populations depend on stellar mass and
this dependence is larger than what has been seen for local density. Even
the two blue populations show large differences for low and high stellar
mass. As a function of stellar mass we can now see differences in the
field and cluster distributions when inspecting each of the four galaxy
populations. However, as before the largest variation is seen for the
highest mass bin (solid curves) of the ``red discs'', indicating that as
these galaxies move from the field to clusters (and increase in mass)
their star formation rate (SFR) decreases.

From these two figures we can infer that both the local and global
environment affect the sSFR, but the former may be more important.
However, the most effective parameter to all the four galaxy populations
is the stellar mass, both in the field and clusters.

%\end{figure}

We have also performed a similar analysis considering the D$_n$(4000) parameter
that traces evolved stellar populations ($> 1$ Gyr, \citealt{kau04, cha15}).
The D$_n$(4000) distributions show similar behaviour to the sSFR ones,
reinforcing that stellar mass is the most effective parameter affecting all
four populations, with the red discs being the most sensitive one.
Other parameters, such as H${\delta}$ (traces the presence of young stellar
populations, $< 1$ Gyr), L$_{dust}$ and $Z$ lead us to similar conclusions.

\subsection{From Blue to Red Discs Across Different Environments}

The study of the variation of galaxy populations (traced by morphological
or star-formation indicators) with environment can be used to constrain
galaxy formation and evolution, and in particular, to understand the
mechanisms responsible for quenching star formation. For instance,
the variation of the star-forming or blue fractions with X-ray luminosity
(analogous to the relations shown in $\S$3) can be used as evidence in
favor (or not) of the {\it ram-pressure stripping} effect
(see \citealt{lop14, rob16}).

We can also use relations between different galaxy parameters for that
purpose. In the current work, if we assume the
``red discs'' to originate in the blue cloud, being a transition
population to the red sequence, it is interesting to compare the
location of blue and red discs in different parameter spaces.

Some of these correlations ({\it {e.g.}}, $Z-$M$_*$) are also useful to
probe different scenarios for transforming galaxies and quench their
star formation. As recently proposed by \citet{pen15}
the relation between stellar metallicity and stellar mass is an important
tool to probe which mechanism is more relevant to quench galaxies. What is
actually used is the metallicity difference between star-forming and passive
populations. Rapid quenching ({\it strong outflows or
ram-pressure stripping}) with sudden removal of the gas reservoir of the
galaxy would imply in small metallicity differences between different
galaxy populations. In the slow quenching scenario, through
{\it strangulation}, the galaxy would still form stars for a long period
with its surviving gas content. Until it is exhausted
metallicity would still grow,
implying in large differences between passive and star-forming objects
(see \citealt{pen15, rob16}).

\begin{figure*}
%\begin{figure}
\begin{center}
\leavevmode
\includegraphics[width=7.0in]{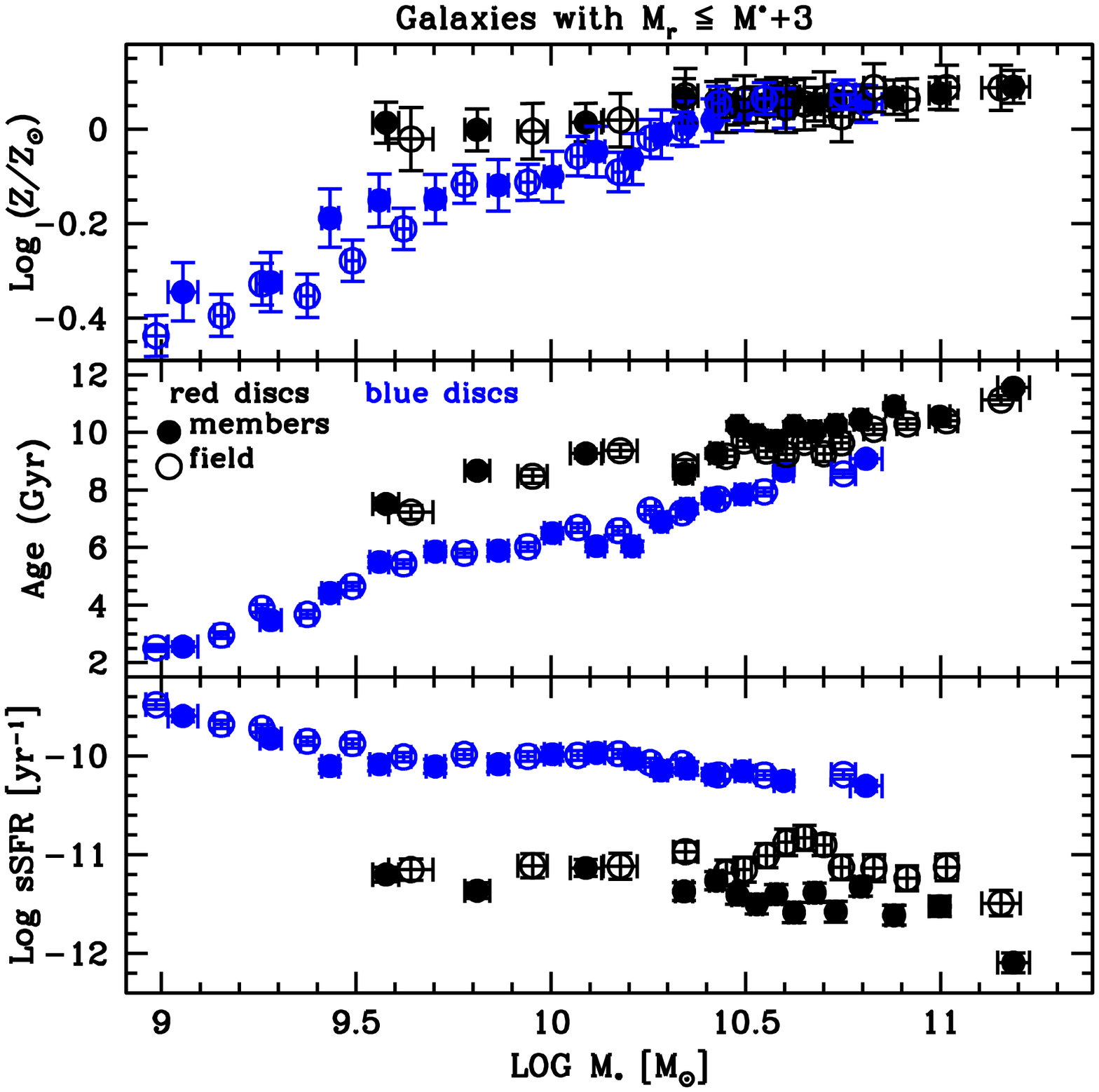}
\end{center}
\caption{The relations between stellar metallicity ($Z$, top panel), the age of
  the stellar population (middle), and the sSFR (bottom) {\it vs}
  stellar mass, for
  the ``blue discs'' (blue points) and ``red discs'' (black points). Field
  galaxies are shown as open symbols and cluster objects as filled symbols.
  The error bars indicate the 1$\sigma$ standard error on the biweight
  location estimate. We detect a clear metallicity difference between
  ``red discs'' and ``blue discs'', from which we estimate a quenching time
  scale of $\sim$ 2$-$3 Gyr.}
\label{fig:ZagesSFR_mass}
\end{figure*}

With that in mind we plot in Fig.~\ref{fig:ZagesSFR_mass} the relations
between stellar metallicity ($Z$), the age of the stellar population, and
the sSFR {\it vs} stellar mass, for the ``blue discs'' (blue points) and
``red discs'' (black points). We also discriminate between field (open
symbols) and cluster (filled symbols) results. The error bars indicate
the 1$\sigma$ standard error on the biweight location estimate. In
the top panel we display the relation with $Z$, while age is shown in
the middle and sSFR in the bottom panel. We can see a clear difference in
the top panel results for red and blue discs,
especially for lower masses. Note also
that ``blue discs'' extend to a lower mass regime, while the opposite is
true for the ``red discs''.  We also detect a small difference between
field and cluster for low mass ``blue discs'' (Log M$_*  \sim $ 9.5, in the
top panel) and for massive ``red discs'' (Log M$_*  > $ 10.5, in the bottom
panel).

The metallicity difference between ``red discs'' and ``blue discs'', in
the top panel of Fig.~\ref{fig:ZagesSFR_mass}, is seen both in
the field and in clusters. An interesting feature is the fact the
differences are slightly larger for field galaxies than cluster objects. The
metallicity difference at Log M$_*$ = 9.5 is $\sim 0.22$ dex for the field
and $\sim 0.14$ dex for cluster galaxies. These values decrease steeply for
Log M$_* \sim$ 10. Comparing these results to the right panel of Figure 2
from \citet{pen15} we would estimate that field ``red discs'' are observed
around 2$-$3 Gyr after quenching due to strangulation, with the time scale
for cluster galaxies being $\sim $ 2 Gyr. The different time scale between
field and cluster galaxies could indicate the importance of the cluster
environment to accelerate galaxy quenching. Mechanisms such as
{\it ram-pressure stripping} (common in the cluster cores) could add up
to the strangulation process, resulting in a shorter time scale for
galaxies to halt star formation. In the middle panel of
Fig.~\ref{fig:ZagesSFR_mass} we also detect an approximately constant
age difference, with stellar mass, of $\sim $ 2 Gyr between the two
populations. However, a proper comparison of the time scales
inferred from the metallicity and age differences would require a detailed
analysis of the SFR, which is beyond the scope of this work.

It is important to emphasize the comparison we make is not between
a star-forming galaxy population and a truly passive population, as we still
have residual star formation among the ``red discs''. If we would isolate
only the SF ``blue discs'' (from the BPT diagram) and the passive
``red discs'' the differences we obtain would be larger, closer to those
from \citet{pen15} and \citet{rob16}.
However, we decided to show the results as in
Fig.~\ref{fig:ZagesSFR_mass} as we assume ``red discs'' to be a transition
population between ``blue discs'' and ``red bulges''. It is encouraging
to detect such metallicity differences between the populations we show.

%\end{figure}

\begin{figure*}
%\begin{figure}
\begin{center}
\leavevmode
\includegraphics[width=7.0in]{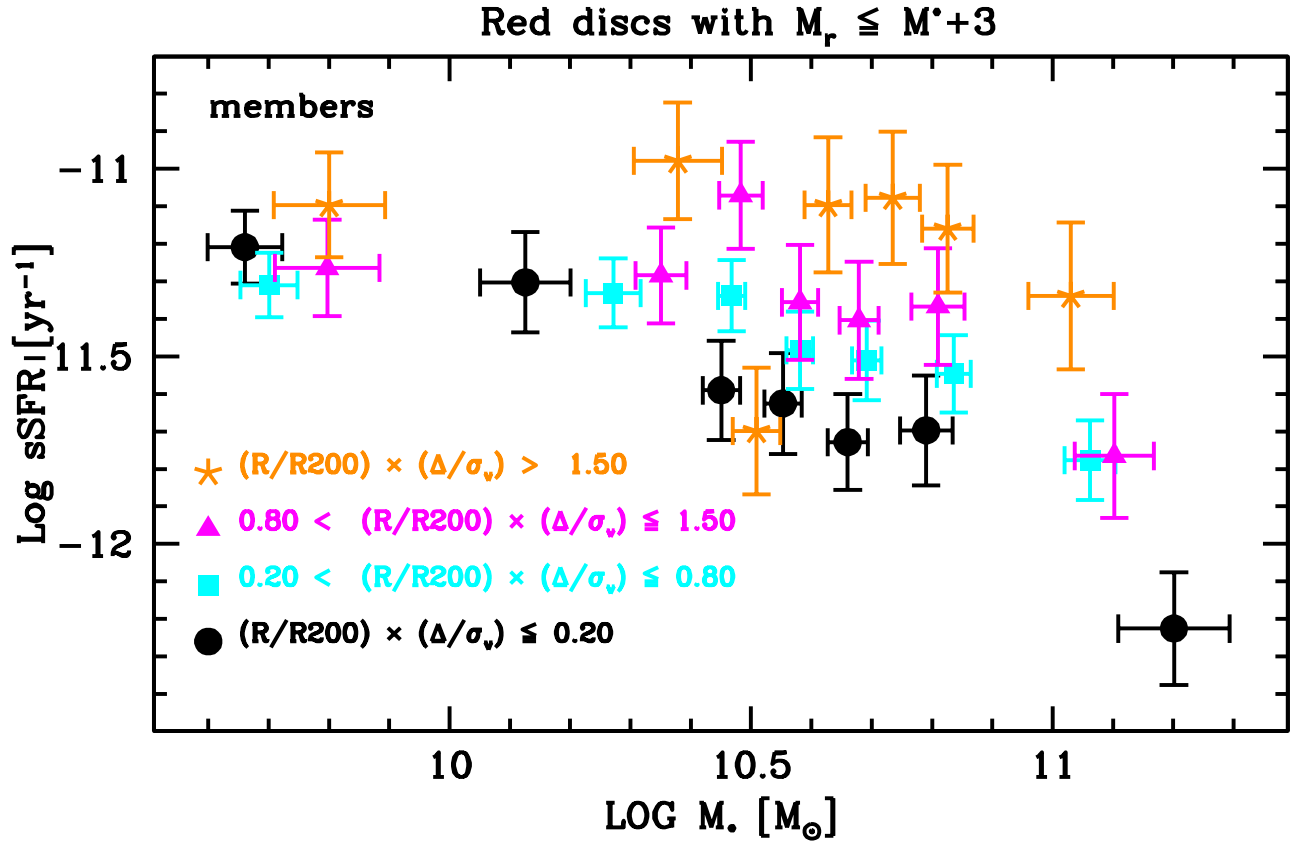}
\end{center}
\caption{The sSFR$-$M$_*$ relation for
``red discs'', dividing the results in four ranges of the phase-space
  parameter (R/R200) $\times $ ($\Delta_v/\sigma_v$). We show in
  dark orange, magenta, cyan and black the results for the intervals
  (R/R200) $\times $ ($\Delta_v/\sigma_v) > $ 1.5,
  0.8 $ < $ (R/R200) $\times $ ($\Delta_v/\sigma_v) \le $ 1.5,
  0.2 $ < $ (R/R200) $\times $ ($\Delta_v/\sigma_v) \le $ 0.8, and
  (R/R200) $\times $ ($\Delta_v/\sigma_v) \le $ 0.2, respectively.
  The error bars indicate the 1$\sigma$ standard error on the biweight
  location estimate. ``Red discs'' decrease their sSFR as they move into
  clusters and grow in mass.}
\label{fig:sSFR_mass_arov}
\end{figure*}

The difference between field and cluster for massive ``red discs''
(Log M$_*  > $ 10.5), seen in the bottom panel of
Fig.~\ref{fig:ZagesSFR_mass}, motivated us to investigate
further the dependence of
the ``red disc'' population in the sSFR$-$M$_*$ plane for different
environments. First we consider only cluster members, in order to verify
how these galaxies are affected when being accreted by clusters. To do
so we show in Fig.~\ref{fig:sSFR_mass_arov} the sSFR$-$M$_*$ relation for
``red discs'', dividing the results in four ranges of the phase-space
parameter (R/R200) $\times $ ($\Delta_v/\sigma_v$) (see \citealt{nob13}).
We consider galaxies that are probably infalling,
or being recently accreted, galaxies
in an intermediate bin, and a central bin with virialized objects (accreted
earlier in the cluster formation). These regions are shown in dark orange,
magenta, cyan and black in Fig.~\ref{fig:sSFR_mass_arov}, for the intervals
(R/R200) $\times $ ($\Delta_v/\sigma_v) > $ 1.5,
0.8 $ < $ (R/R200) $\times $ ($\Delta_v/\sigma_v) \le $ 1.5,
0.2 $ < $ (R/R200) $\times $ ($\Delta_v/\sigma_v) \le $ 0.8, and
(R/R200) $\times $ ($\Delta_v/\sigma_v) \le $ 0.2, respectively. Note these
regions are not precisely defined and serve for guidance only.

From Fig.~\ref{fig:sSFR_mass_arov} we can see a clear decrease in sSFR as we
go from high to low values of the phase-space parameter
(R/R200) $\times $ ($\Delta_v/\sigma_v$), indicating that ``red disc''
galaxies halt the star formation when moving into the clusters.
However, the strong dependence to stellar mass is still seen. For instance,
for the most central bin (black points) the sSFR decreases from
Log sSFR $\sim $ -11.2 to Log sSFR $\sim $ -12.2, at
9.7 $<$ Log M$_*  < $ 11.2.

%\end{figure}

\begin{figure*}
%\begin{figure}
\begin{center}
\leavevmode
\includegraphics[width=7.0in]{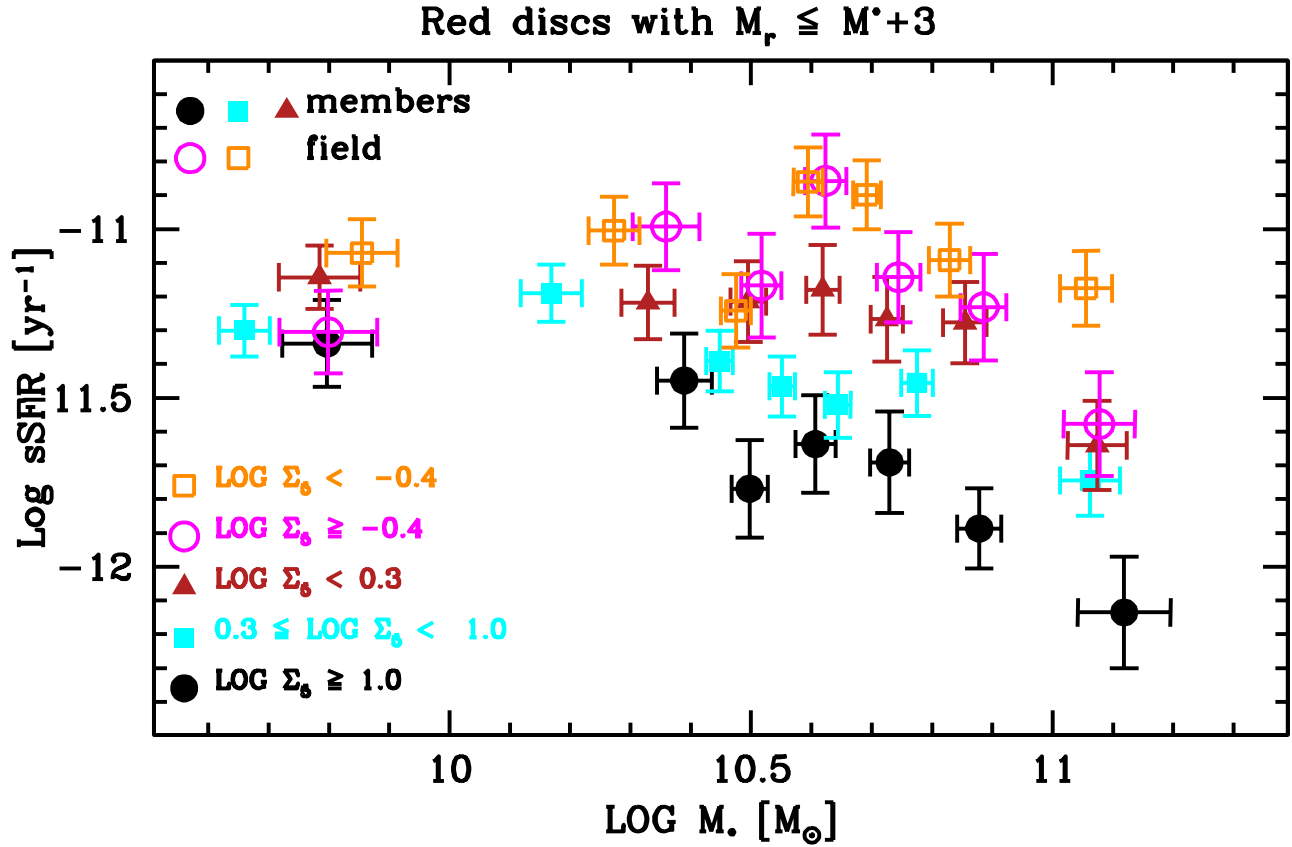}
\end{center}
\caption{Analogous to Fig.~\ref{fig:sSFR_mass_arov}, but now
  also showing field results, and dividing the ``red discs'' in local
  galaxy density intervals, two for the field and three for cluster
  galaxies. In the field ``red discs'' are shown by open symbols, in
  dark orange and magenta for the intervals Log $\Sigma_5 <$ -0.4 and
  Log $\Sigma_5 \ge$ -0.4, respectively. Member galaxies are displayed by
  filled symbols, in brown, cyan and black, for the intervals
  Log $\Sigma_5 <$ 0.3, 0.3 $\le$ Log $\Sigma_5 <$ 1.0 and
  Log $\Sigma_5 \ge$ 1.0, respectively. The error bars indicate the
  1$\sigma$ standard error on the biweight location estimate. Except for the
  most massive, field ``red discs'' are not affected by local environment.
  On the contrary, for cluster members the sSFR is sensitive to local
  density and stellar mass for objects more massive than Log M$_* =$ 10.4.}
\label{fig:sSFR_mass_dens}
\end{figure*}

Fig.~\ref{fig:sSFR_mass_dens} is analogous to Fig.~\ref{fig:sSFR_mass_arov},
but now we also show field results, and divide the ``red discs'' in local
galaxy density intervals, two for the field and three for cluster galaxies.
The density ranges adopted are indicated in the figure. Open symbols are
used for field results and filled symbols for cluster galaxies.
This plot reinforces the conclusion from the previous figure, indicating that
cluster ``red discs'' decrease their sSFR when reaching higher local density
values (or smaller values of (R/R200) $\times $ ($\Delta_v/\sigma_v$)).
However, now comparing to the field ``red discs'' we can see this population
(open points) assumes a nearly constant value (and higher than for
clusters), except for the most massive bin. Hence, field ``red discs'' are
not affected by local environment (except the most massive,
Log M$_* >$ 10.8). For cluster members the sSFR is sensitive to
local density and stellar mass for objects more massive than
Log M$_* =$ 10.4. The variation with stellar mass is also more
pronounced for objects in environments with Log $\Sigma_5 >$ 0.3 (cyan and
black points),
which is roughly equivalent to
(R/R200) $\times $ ($\Delta_v/\sigma_v) \le $ 0.8 (from
Fig.~\ref{fig:sSFR_mass_arov}).

\begin{figure*}
%\begin{figure}
\begin{center}
\leavevmode
\includegraphics[width=7.0in]{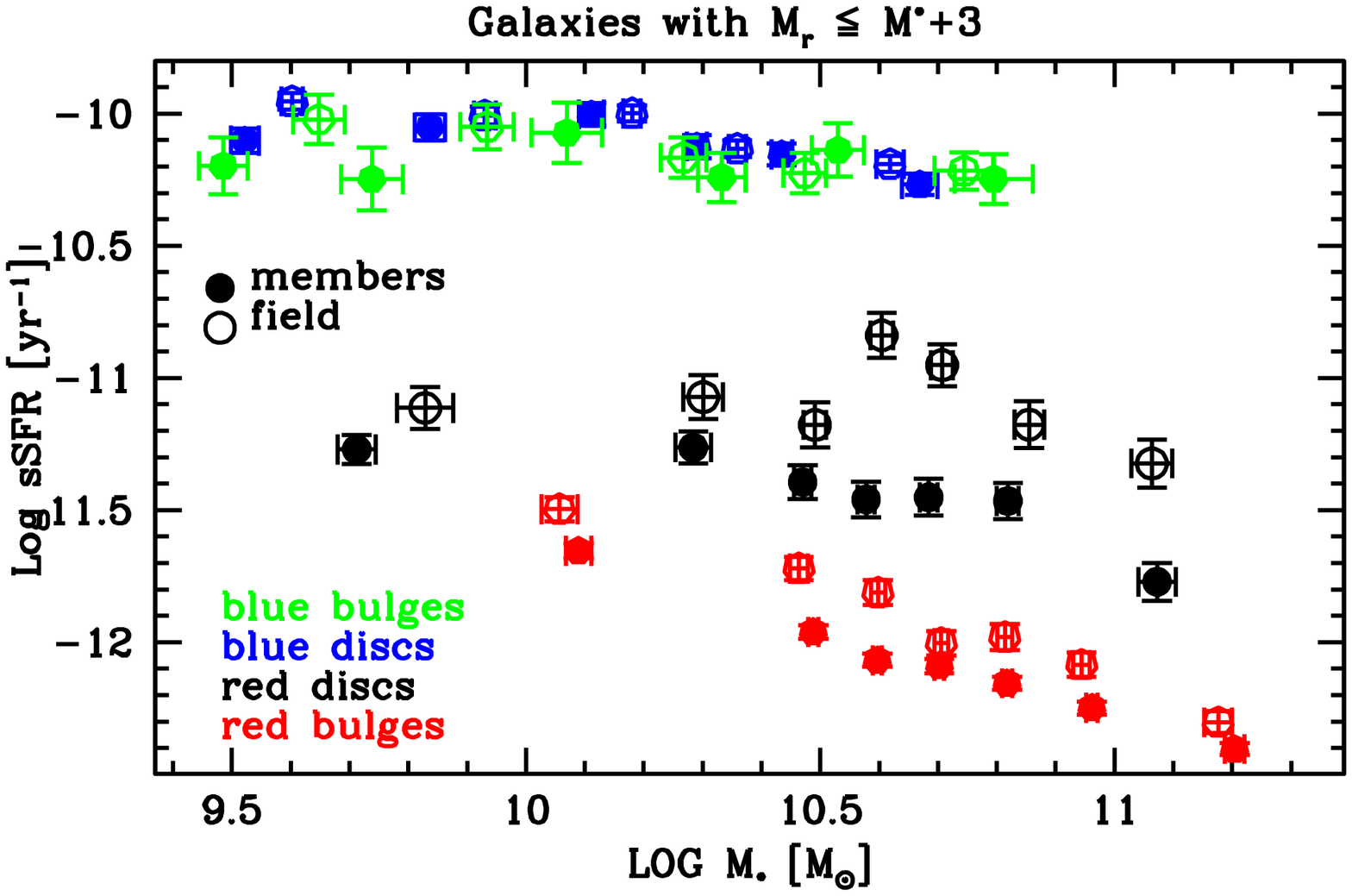}
\end{center}
\caption{Analogous to Fig.~\ref{fig:sSFR_mass_dens}, but now
  also showing the other three populations (``blue discs'', ``blue bulges''
  and ``red bulges''), and not splitting the data in density bins. For each
  galaxy population field results are shown by open symbols, while members
  are displayed by filled symbols. The error bars indicate the 1$\sigma$
  standard error on the biweight location estimate. In the stellar mass
  range of this figure only the red galaxies decrease the sSFR as they grow
  in mass. Those objects are also the only ones to display differences
  between field and cluster results, especially the ``red discs''.}
\label{fig:sSFR_mass_dens2}
\end{figure*}

Fig.~\ref{fig:sSFR_mass_dens2} is similar to Fig.~\ref{fig:sSFR_mass_dens},
but we do not split the data in density bins. We only compare field (open
symbols) and cluster galaxies (filled symbols), but now besides showing the
``red discs'' (black points) we also exhibit the other three populations;
``blue discs'', ``blue bulges'' and ``red bulges'', in blue, green and red,
respectively. Comparing to the ``blue discs''
and ``red bulges'' we reinforce the suggestion that ``red discs'' represent
a transition population between the other two. As ``blue discs'' grow in
mass they decrease their specific star formation rate. As they move
into clusters, where other mechanisms are common (such as
{\it ram-pressure stripping}), this effect is probably enhanced.
Hence, we could suggest the transformation of ``blue discs'' into
``red discs'' to be the combination of the stellar mass growth and
environmental influence. Another interesting feature on this figure is the
fact that only the red galaxies display differences between field and cluster
results, especially the ``red discs''.

%\end{figure}

%\end{figure}

\subsection{The Star Formation Histories (SFH)}

Another way to compare the two transitional galaxy populations (``red discs''
and ``blue bulges'') to the other galaxies (``blue discs'' and ``red bulges'')
is through the use of their star formation history estimates. The 
STARLIGHT code derives the amount of star formation for a galaxy in several
age bins. The current stellar mass of each galaxy is the result of adding
the stellar mass formed on each age bin. We express the star formation history
(SFH) of each galaxy as the fraction of stars formed on each age bin. Here we
consider four wide age bins (as done by \citealt{toj13}). Note the
star formation fractions add up to unity over all cosmic history.

\begin{figure*}
%\begin{figure}
\begin{center}
\leavevmode
\includegraphics[width=7.0in]{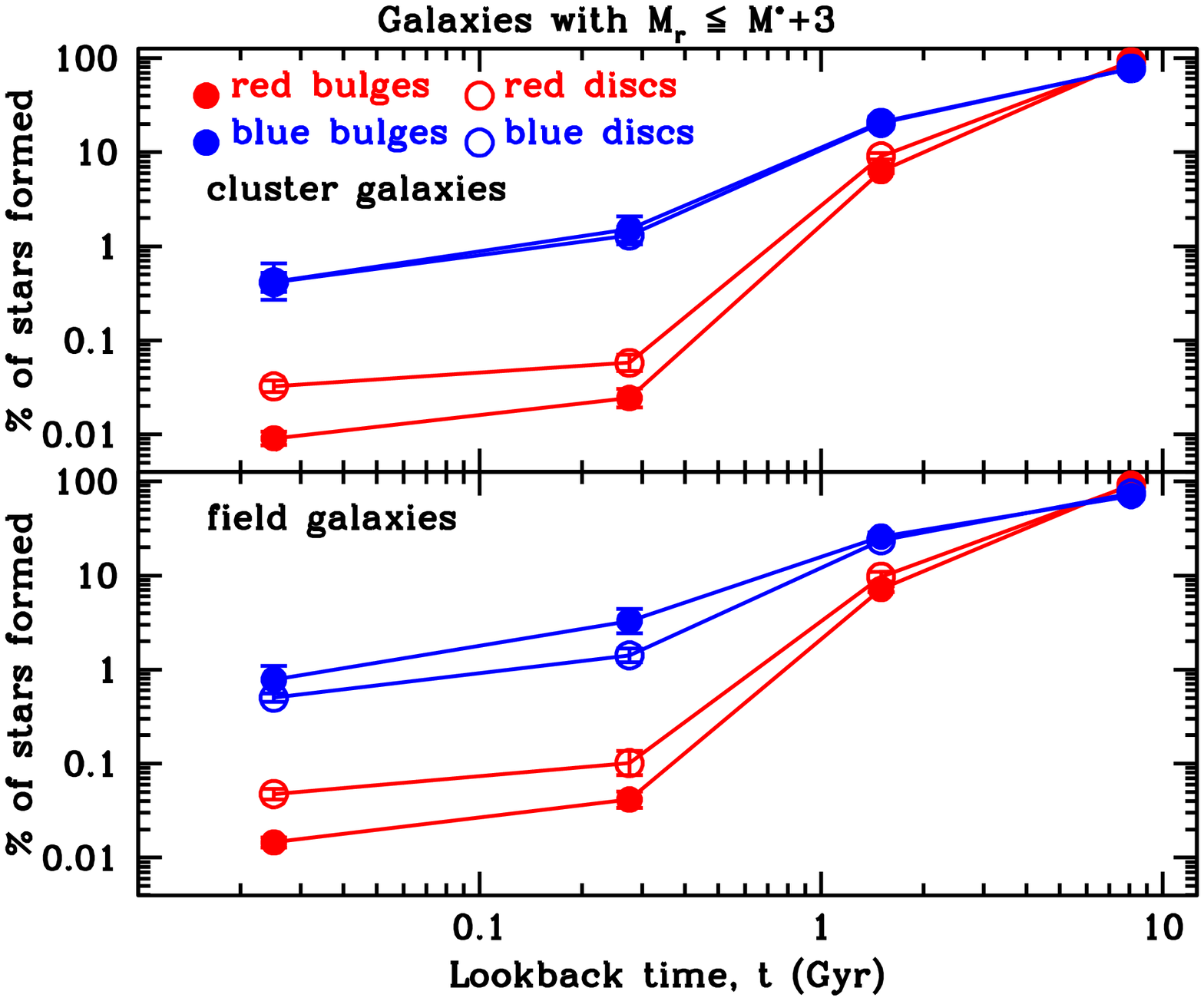}
\end{center}
\caption{The star formation history (average star formation fraction as a
  function of lookback time) for the four galaxy populations in the current
  work. Red galaxies are displayed in red, and blue galaxies in blue. Bulges
  are shown as filled symbols, while discs are exhibited as open symbols. The
  top panel shows the results for cluster galaxies, while the field is in
  the bottom. The four age intervals correspond to 0$-$0.05 Gyr,
  0.05$-$0.5 Gyr, 0.5$-$2.5 Gyr, and $>$ 2.5 Gyr. The error bars
  indicate the 1$\sigma$ standard error on the biweight location estimate.
  The ``red discs'' form $\sim $ 4 times more stars than ``red bulges'' in
  the recent Universe ($< $ 100 Myr). The SFHs of red and blue discs are
  significantly different for at least $\sim $ 2.5 Gyr, corroborating the
  quenching time scale estimated after
  Figs.~\ref{fig:ZagesSFR_mass}$-$\ref{fig:sSFR_mass_dens2}.}
\label{fig:sff_lookbt}
\end{figure*}

Fig.~\ref{fig:sff_lookbt} shows the average
star formation fractions (SFFs) as a
function of lookback time for the four galaxy populations (``red discs'',
``blue bulges'', ``blue discs'' and ``red bulges''). Bulges are shown as
filled symbols, and discs as open symbols. Blue galaxies are in blue and red
objects in red. We also split cluster and field results, with the former in
the top panel and the latter in the bottom. The four age intervals correspond
to 0$-$0.05 Gyr, 0.05$-$0.5 Gyr, 0.5$-$2.5 Gyr, and $>$ 2.5 Gyr.

%\end{figure}

From Fig.~\ref{fig:sff_lookbt} we can see the recent star formation of
``red discs'' is smaller than the ``blue discs'' (by 10 to 12 times, if in the
field or clusters), but is larger than the ``red bulges''. The ``red disc''
population forms $\sim $ 4 times more stars than ``red bulges'' in the recent
Universe ($< $ 100 Myr). Differences in the SFHs of red and blue discs are
seen in all age bins, especially the most three recent. That indicates their
SFHs are significantly different for at least $\sim $ 2.5 Gyr. This result
corroborates the quenching time scale we had estimated for the ``blue discs''
from Figs.~\ref{fig:ZagesSFR_mass}$-$\ref{fig:sSFR_mass_dens2}.
We also find field and cluster results to be very
similar, except for the fact the field SFFs are slightly
higher, especially for the two youngest bins. Finally, blue galaxies
(bulges and discs) have very similar SFHs, except for the field, where
``blue discs'' form less stars in the two youngest bins. Note that our results
are not directly comparable to the ones from \citet{toj13} as they
allow for more galaxy populations, dividing the spirals in blue and late types.
They also restrict their analysis to massive galaxies (Log M$_*  > $ 10.7).
However, we verified that there is no difference in our conclusions if we
restrict the analysis to massive galaxies.

\subsection{The Nature of Blue Bulges}

The location of ``blue discs'' is very similar to the ``blue bulges'' in
Fig.~\ref{fig:sSFR_mass_dens2}. The two populations show a decrease in the
sSFR with stellar mass. Note the decrease would be even larger if we were
showing the less massive objects (Log M$_*  \le $ 9.5), as can be seen for
``blue discs'' in the bottom panel of Fig.~\ref{fig:ZagesSFR_mass}. The two
populations also show little difference between field and cluster results.
The SFHs displayed in Fig.~\ref{fig:sff_lookbt} are also very similar for the
two populations. Hence, despite the fact ``blue discs'' and ``blue bulges''
have different morphologies they do show similar SFR properties, but
slight different distributions of age and metallicity. We then decided to
investigate if part of the ongoing star formation in the ``blue bulges''
could be the result of mergers.

We do so as the visual inspection of the ``blue bulges'' indicates some of
them show signs of interaction. We investigated this further estimating the
asymmetry of these galaxies, and also comparing them to the ``red bulges''.
To derive asymmetry for these galaxies we used the public software PyCA
\citep{men06}. PyCA is a Python software designed to compute
asymmetry (A) and concentration (C) from SExtractor products. Note the
definition of concentration in PyCA is different from the SDSS, which we
adopt for this work. More details on this software can be found in
\citet{men06}. We computed the ``A'' parameter for all
``blue bulges'' and for comparison for a subset of ``red bulges'' (900
galaxies). The software is applied to the $r$-band images, which are deep
enough and not too affected by local star formation.

\begin{figure*}
%\begin{figure}
\begin{center}
\leavevmode
\includegraphics[width=7.0in]{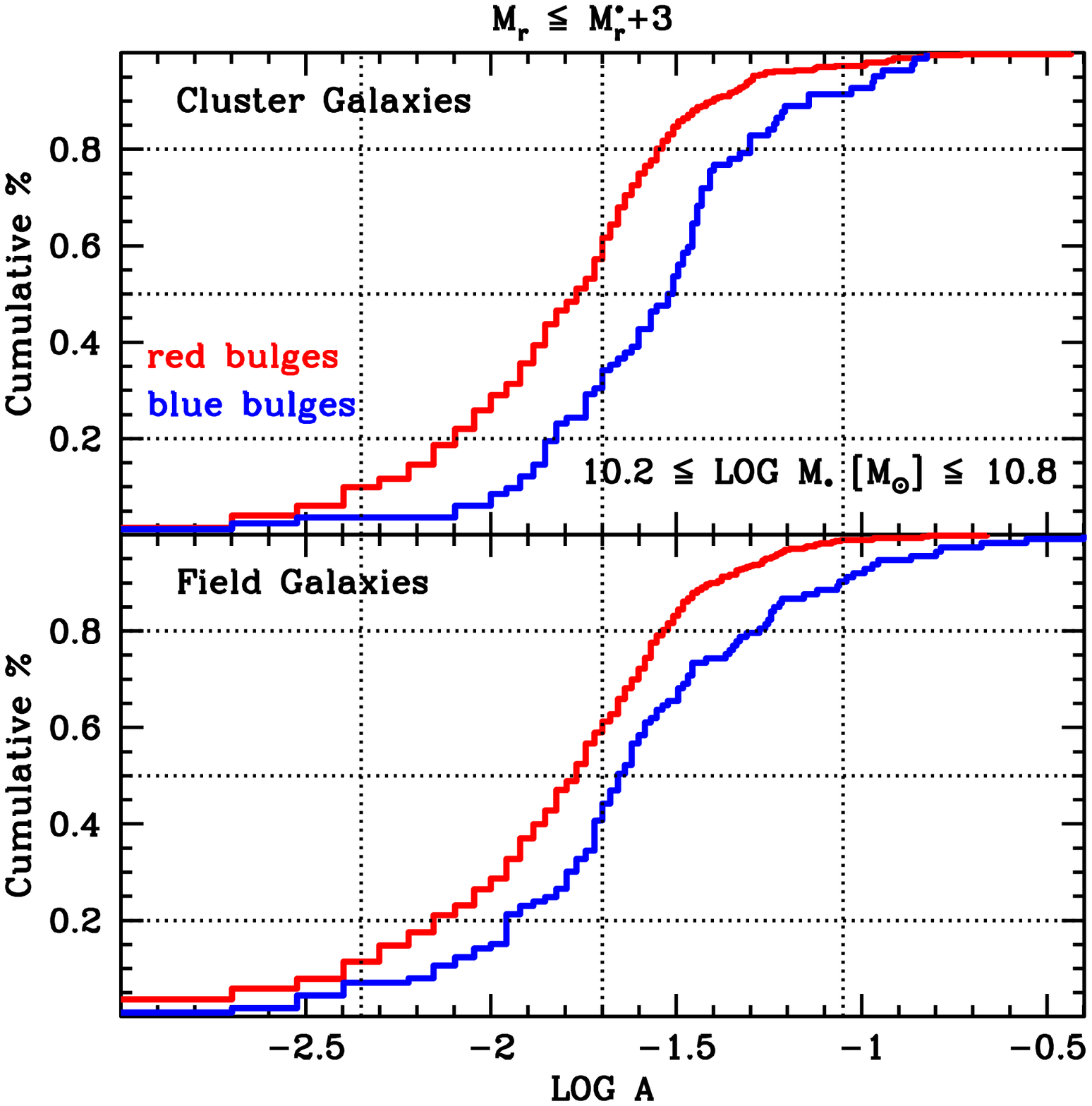}
\end{center}
\caption{The cumulative distribution of the asymmetry (A) parameter for the
  red and blue bulges at 10.2 $\le$ Log M$_*  \le $ 10.8. In the top
  the comparison considers cluster galaxies, while in the bottom only field
  galaxies. Even for this small stellar mass range we clearly see the
  ``blue bulge'' population show larger values of the asymmetry (A) parameter.}
\label{fig:A_bb_rb}
\end{figure*}

We compared the asymmetry cumulative distributions of all ``blue bulges''
and the subset of ``red bulges'', finding the former to be shifted to higher
A values than the latter. As the two populations span different stellar mass
ranges we then performed the comparison only for galaxies in a common mass
range (10.2 $\le$ Log M$_*  \le $ 10.8), as displayed in
Fig.~\ref{fig:A_bb_rb}, in the top for cluster and in the bottom panel
for field galaxies. Even for this small stellar mass range the differences
between the asymmetry distributions for the two populations are easily seen.
The comparison between cluster and field results in Fig.~\ref{fig:A_bb_rb}
also reveals no difference for the ``red bulges'', but slightly higher
asymmetry values for the ``blue bulge'' cluster population.
That reinforces the discussion related to
Fig.~\ref{fig:age_metl_mass} ($\S$ 4), where we argue that ``blue bulges''
in the field and in clusters may be composed of different subtypes of
galaxies. That could be one reason explaining higher $Z$ values for 
``blue bulges'' in clusters.

%\end{figure}

From the visual inspection of extreme cases (low and high asymmetry) of
red and blue bulges we find the following. In many cases the ``blue bulges''
show clear signs of strong interaction, indicating wet mergers. However,
in some cases they look like very asymmetric
low mass spiral galaxies dominated by
a bulge. On the other hand, for the ``red bulges'' with high A values, when
the asymmetry is clearly distinguishable they only display signs of dry
mergers.

For the galaxies with low asymmetry, both populations (``blue bulges''
and ``red bulges'') look like ``spheroidal'' or bulge dominated systems.
In particular, the ``blue bulges'' are small and often display a starburst
spectrum. Hence, we can say that from low to high asymmetry the
``blue bulges'' can be simple spheroids,
or bulge dominated low mass spirals, or
wet mergers. However, it is important to stress that this visual inspection
we perform is not the main goal of the current work (as we discussed in
$\S$ 2.6). In particular, the ``blue bulges'' are generally faint low mass
systems, making their visual classification harder. When we say some of them
look like very asymmetric spiral galaxies, that does not mean we are
referring to large massive early spirals.

\begin{figure*}
%\begin{figure}
\begin{center}
\leavevmode
\includegraphics[width=7.0in]{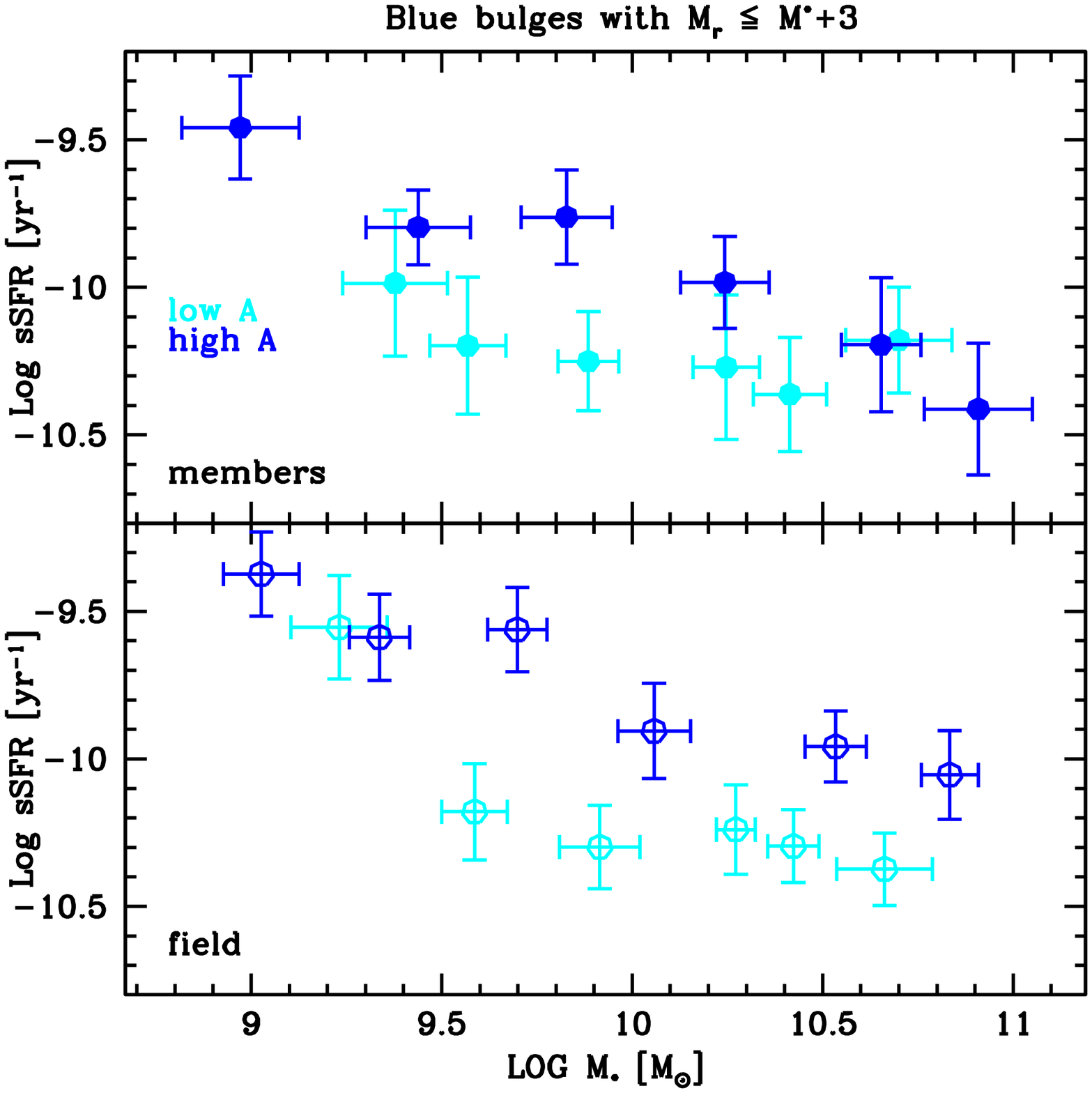}
\end{center}
\caption{The sSFR {\it vs} M$_*$ plane for ``blue bulges'' classified as
  low (cyan points) and high (blue points) asymmetry. In the top
  the comparison considers cluster galaxies, while in the bottom only field
  galaxies. At 9.5 $<$ Log M$_* < $ 10.5 there is a significant difference
  in the sSFR of low and high A ``blue bulges'', with highly asymmetrical
  galaxies having higher sSFR values.}
\label{fig:ssfr_mstar_lhA_bb}
\end{figure*}

Considering the asymmetry value we also compared the specific star formation
rate {\it vs} stellar mass plane of low and high A ``blue bulges''. That is
displayed in Fig.~\ref{fig:ssfr_mstar_lhA_bb}. We can see a significant
difference in the sSFR of low and high A ``blue bulges'' at 
9.5 $<$ Log M$_* < $ 10.5 (note that does not happen at all masses),
with high A galaxies displaying higher sSFR
values. That indicates these mergers contribute to make part of
the the ``blue bulge'' population more active than the rest. A similar
comparison, but using H${\delta}$ (instead of sSFR), also reveals differences
between low and high A ``blue bulges'', indicating the presence of
young stellar populations ($< 1$ Gyr) among the most asymmetrical
``blue bulges''. A similar plot using metallicity ($Z-$M$_*$) shows
no significant difference, reinforcing the SFR enhancement is recent.

%\end{figure}

\section{Discussion}

Several different works in the literature aim to investigate transitional
galaxies and to assess the time scale for quenching star formation
\citep{wol09, muz14, vul15, hai13, hai15, toj13, rob16}.
Many results point to a
slow quenching scenario on which processes like strangulation are
predominant \citep{pen15, rob16, mai16}. On top of that the importance of the
cluster environment has been strengthened, especially by studies based
on the investigation of galaxy properties in the phase-space
\citep{nob13, nob16, hai13, hai15, mai16}.

\citet{hai13} find the specific SFRs of massive central star-forming
galaxies (within R200) to be much lower than the field counterpart.
They take that as an indication of slow quenching (timescales of
0.7$-$2.0 Gyr) of massive SF galaxies once arriving in clusters.
\citet{hai15} verify the surface density of SF to have a steep
decline with radius, implying that recently accreted spirals will maintain
star formation for 2$-$3 Gyr. They also compare the phase-space diagram of
SF galaxies with results from the Millennium simulation, concluding the
quenching time scale to be $\sim $ 1.7 Gyr.

Although we are mainly interested in transitional galaxies (in particular
``red discs'' as an intermediate population between the blue cloud and red
sequence), our conclusions are in good agreement to those above. From 
Figs.~\ref{fig:ZagesSFR_mass}$-$\ref{fig:sff_lookbt} we have different
indications that the star formation quenching is slow ($\sim $ 2 Gyr)
and depends on stellar mass, but also on environment. The metallicity
difference as a function of stellar mass, between ``red discs''
and ``blue discs'', is consistent to a quenching time scale of 2$-$3 Gyr
for field and cluster galaxies.

From Fig.~\ref{fig:sff_lookbt} we conclude ``red discs'' have
residual star formation compared to the ``blue discs'', but still forms
$\sim $ 4 times more stars than ``red bulges'' in recent times ($< $ 100 Myr).
The comparison of the SFHs of ``red discs'' and ``blue discs'' lead us
to the conclusion their SFH are significantly different
for at least $\sim $ 2.5 Gyr.
All these facts, plus the distributions of SFR, $12 \mu m$ luminosity,
L$_{dust}$ and sSFR, indicate the red and blue discs are distinct populations,
with the former possibly representing the transition to ``red bulges''. This
transition is slow, with quenching time scales of $> $ 2 Gyr. Morphological
transformations would happen even in a longer time scale, which is also
in good agreement to a scenario requiring pre-processing in groups
\citep{hai15, lop14}. In the pre-processing scenario galaxy transformations
could be divided in two steps \citep{lac13}, where star formation is quenched
in the group scale, but morphological transformation is a separate process,
occurring in clusters. First, star formation is halted in discs residing
in relatively low-density environments. Secondly, a morphological
transformation from disc to bulge-dominated systems occur at higher densities.

Figs.~\ref{fig:sSFR_mass_arov}$-$\ref{fig:sSFR_mass_dens2} clearly
point to the importance of the cluster environment to decrease the
sSFR of the transitional galaxies called ``red discs''.
In Figs.~\ref{fig:sSFR_mass_dens}$-$\ref{fig:sSFR_mass_dens2} we can
also detect the difference between field and cluster results for the
``red disc'' population. The global picture is that ``blue discs''
decrease their specific star formation rate as they grow in mass and as
they infall into galaxy clusters. We can see (Fig.~\ref{fig:sSFR_mass_dens})
a clear difference between field (dark orange and magenta points) and
cluster (brown, cyan and black points) results for the ``red discs''. On
top of that we detect a significant dependence with local density for
the massive cluster ``red discs'' (Log M$_*  > $ 10.5).

On Fig.~\ref{fig:sSFR_mass_dens2} we can also see the variation of the sSFR
with stellar mass for the ``red bulges'' (varying from Log sSFR
$\sim $ -11.5 to Log sSFR $\sim $ -12.5, at
10.0 $<$ Log M$_*  < $ 11.5). It is
also possible to notice the difference between field (open symbols) and
cluster results (filled points) for the ``red bulges''. That is not true
for the star-forming blue galaxies (``blue discs'' and ``blue bulges'').
We can not detect a significant difference between field and cluster results,
and only a small variation of sSFR with stellar mass. A large variation exists
only for lower mass galaxies (Log M$_*  < $ 9.5), as can be seen in
the bottom panel of Fig.~\ref{fig:ZagesSFR_mass}. These results are in good
agreement to what is seen in \citet{hai13}. They find a significant
difference in the sSFR {\it vs} M$_*$ relation for field and cluster
massive star-forming galaxies, but only when using members within R200.
If the cluster sample is from the infall region,
field and cluster results agree. We do not make an environmental distinction
within clusters for
the blue galaxies (discs and bulges); so that the good agreement we see
to the field is probably due to predominance of these SF blue galaxies in
the lower density infall regions of clusters (closer in density to the field
values).

This similarity between ``blue discs'' and ``blue bulges'' seen on the
sSFR$-$M$_*$ relation (Fig.~\ref{fig:sSFR_mass_dens2}) is also detected
in several other properties (see, for instance, Figs.~\ref{fig:bptclass_sfr},
~\ref{fig:rad_tc_sig5},~\ref{fig:sSFR_mass_dens},~\ref{fig:sff_lookbt}).
Although these two populations seem to have similar SFR and dust luminosity,
they display slightly different distributions of age, metallicity and stellar
mass for bright galaxies (Fig.~\ref{fig:age_metl_mass}). They also display
different dependence on environment for the sSFR distributions seen in
Fig.~\ref{fig:ssfr_densbins}.
However, it is important to stress that blue bulges and discs similarities
are much more pronounced than their differences. Except for morphology
``blue bulges'' are much closer to ``blue discs'' than to ``red bulges''.
\citet{vul15} suggest that blue star-forming early-types (BSF) could
be the result of a morphological transformation happening before the
star formation is halted, with the stellar disc being removed entirely or at
least in part. On the other hand, \citet{toj13} argue that the
differences in the dust content between blue ellipticals and spirals is
an indication the former are not descendants of the blue spirals. Another
possibility is that these blue early-type galaxies be the result of
rejuvenation, through the merger with a star forming galaxy.
\citet{sal10} suggest this rejuvenation scenario is plausible
from the analysis of
optically quiescent early-type galaxies with strong UV excess.
\citet{kan09} argue that blue E/S0 individual galaxies may be
evolving either up to red sequence or down into the blue cloud. They also
argue those galaxies represent a transitional class. The most massive resemble
major-merger remnants that will end up in the red sequence, while lower mass
objects (M $< $ 3 $\times$ 10$^{10}$ M$_{\odot}$) display signs of disk and/or
pseudobulge building. The case for lower mass galaxies is reinforced by the
results of \citet{wei10}, who investigated low mass blue-sequence
E/S0 galaxies. They argue these low mass galaxies are more common in
low-density field environments where fresh gas infall is possible.
They find evidence that star formation is bursty, involving externally
triggered gas inflows. They also suggest most of these galaxies can grow
stellar disks on relatively short timescales.

As stated above we do find evidence for mergers within the ``blue bulges''
sample, both from their visual inspection and the analysis of asymmetry (see
Figs.~\ref{fig:A_bb_rb} and \ref{fig:ssfr_mstar_lhA_bb}). However, in some
cases the high asymmetry ``blue bulges''
simply resemble asymmetric low mass spiral
galaxies dominated by a bulge. An interesting point is that we find evidence
that high A ``blue bulges'' have an enhanced star formation compared to
the more regular galaxies (see Fig.~\ref{fig:ssfr_mstar_lhA_bb}).
To summarize, from low to high asymmetry
we find ``blue bulges'' that are spheroids, 
bulge dominated low mass spirals, or resemble wet mergers.
Hence, quoting \citet{toj13}, there may not be a
single evolutionary path for the blue
early-type objects. As \citet{kan09} suggest some of these objects
may be going up to the red sequence, while others are going in the opposite
direction towards the blue cloud.

\section{Summary}

In this work we measure the typical environment of transitional galaxy
populations selected solely by photometric parameters (color and
concentration). Those galaxies are called ``red discs'' and ``blue bulges''.
We compare the environments of these galaxies to normal objects from the blue
cloud and red sequence, which we select as ``red bulges'' and ``blue discs''.
Besides comparing environmental related parameters (such as crossing time and
$\Sigma_5$), we also compare physical properties like age and metallicity.
We also used the cumulative distributions of different properties of these
galaxy populations to assess the impact of stellar mass and environment (local
and global). Then we investigate the location of these populations in the
$Z-$M$_*$, Age$-$M$_*$ and sSFR$-$M$_*$ planes, and analyze their SFHs. Doing
so, we suppose possible galaxy evolutionary scenarios and estimate the quenching
time for ``blue discs''. Finally, we estimate  the asymmetry of the
``blue bulges'', trying to assess the impact of mergers on this population.
Our main results are:

\begin{enumerate}

\item At fixed morphology we see the number of ``red discs''
  (``blue bulges'') relative to the number of discs (bulges) vary strongly
  with local density, especially in the cluster environments
  (Figs.~\ref{fig:denbins2},~\ref{fig:denbinsb2},~\ref{fig:radbins2},~\ref{fig:radbinsb2}). The transformation happens mainly within the virial radius.

\item As a function of distance to the cluster center we find the two
  populations to be nearly constant from the outskirts to $R_{200}$.
  Inwards, the relative numbers decrease (increase) for the ``blue bulges''
  (``red discs''), for bright and faint galaxies. Such results give strength
  to the importance of clusters to transform galaxy properties
  (Figs.~\ref{fig:radbins2},~\ref{fig:radbinsb2}).

\item Dividing the sample in stellar mass bins we see this environmental
  variation is most significant for the lower mass bins
  (Log M$_* \le 10.6$), reinforcing that star formation is halted
  first in higher mass objects (Fig.~\ref{fig:den_stmassbins2}).

\item Galaxies of the same color, but different morphologies (discs or
  bulges) show little difference in their typical environment. However, that
  is not true if morphology is fixed, instead of color. Red and blue discs
  are found in very different environments, stressing these two populations
  are indeed different. Red objects have shorter crossing times and are at
  higher densities than blue galaxies (Fig.~\ref{fig:rad_tc_sig5}).
  
\item Although found at similar environments ``red bulges'' have an older
  stellar population and higher metallicities than ``red discs''
  (Fig.~\ref{fig:age_metl_mass}). 

\item The SFR of red and blue discs is very different, and the former shows
  higher values than the passive ``red bulges'' (Fig.~\ref{fig:bptclass_sfr}). 

\item  On what regards dust luminosity ``blue bulges'' have slightly lower
  values than ``blue discs'', which have much larger L$_{dust}$ than
  ``red discs''. Nonetheless, the later still shows higher L$_{dust}$
  values than ``red bulges'' (Fig.~\ref{fig:sl12_dustl_mass}).  

\item From the cumulative distributions of sSFR we see no
  overlap of blue and red populations, and ``red discs'' and ``red bulges''
  are different in all environments. We also find the largest variations
  with local density for the red objects at fixed morphology (mainly for
  the ``red discs'' within clusters). Finally, local galaxy density seems
  to affect more those distributions than the global environment
  (Fig.~\ref{fig:ssfr_densbins}).
  
\item A similar analysis in different stellar mass ranges reveals this
  parameter to be even more important to galaxy properties when
  compared to $\Sigma_5$. The largest variations are again seen for
  ``red discs'', suggesting those galaxies halt star-formation as
  increasing their mass and moving into clusters. Hence, both local
  and global environment matter (local being more important). But
  the most effective parameter to shape all four galaxy populations is
  stellar mass (Fig.~\ref{fig:ssfr_massbins}).

\item The metallicity difference, as function of stellar mass,
  between ``red discs'' and ``blue discs'' is consistent to a slow
  quenching scenario, with a time
  scale of 2$-$3 Gyr for field and cluster galaxies. This time scale is
  corroborated by the analysis of the star formation histories of the
  different galaxy populations. The SFHs of ``red discs'' and ``blue discs''
  should differ for at least $\sim $ 2.5 Gyr
  (Figs.~\ref{fig:ZagesSFR_mass},~\ref{fig:sff_lookbt}).

\item The distribution of ``red discs'' in the sSFR$-$M$_*$ plane for
  different environments reveals this population to gradually change as they
  move into clusters. That is seen in Figs.~\ref{fig:sSFR_mass_arov}, 
  ~\ref{fig:sSFR_mass_dens} and ~\ref{fig:sSFR_mass_dens2}. The
  first figure shows the results in phase-space bins, the second in local
  density bins (also comparing field and cluster ``red discs''). The
  third figure compares field and cluster galaxies, but also displaying
  the other three populations.

\item Fig.~\ref{fig:sSFR_mass_dens2} is also important to reinforce the idea
  that ``red discs'' are the descendants of ``blue discs'', on their way
  to become a passive population.

\item We found that part of the ``blue bulges'' can be explained by wet
  mergers, as indicated by their visual inspection and asymmetry values
  (Fig.~\ref{fig:A_bb_rb}). In particular, the high asymmetry ``blue bulges''
  have larger sSFR and H${\delta}$ values, indicating the presence of
  young stellar populations (Fig.~\ref{fig:ssfr_mstar_lhA_bb}).

\item We do not conclude on a single evolutionary path for the ``blue bulge''
  population, as we find they may be simple spheroids, low mass bulge
  dominated spirals, or resemble wet mergers.

\end{enumerate}

\section*{Acknowledgements}

ALBR thanks for the support of CNPq, grants 306870/2010-
0 and 478753/2010-1. PAAL thanks the support of CNPq, grant
308969/2014-6.

This research has  made use of the SAO/NASA  Astrophysics Data System,
and the NASA/IPAC Extragalactic  Database (NED).  Funding for the SDSS
and  SDSS-II was  provided  by  the Alfred  P.  Sloan Foundation,  the
Participating  Institutions,  the  National  Science  Foundation,  the
U.S.  Department  of  Energy,   the  National  Aeronautics  and  Space
Administration, the  Japanese Monbukagakusho, the  Max Planck Society,
and  the Higher  Education  Funding  Council for  England.  A list  of
participating  institutions can  be obtained  from the  SDSS  Web Site
http://www.sdss.org/.

%%%%%%%%%%%%%%%%%%%%%%%%%%%%%%%%%%%%%%%%%%%%%%%%%%



%\bibliographystyle{mnras}

\begin{thebibliography}{99}

\bibitem[Berlind et al.(2006)]{ber06} Berlind A., Frieman J., Weinberg D. 
et al. 2006, ApJS, 167, 1

\bibitem[\protect\citeauthoryear{Brinchmann et al.}{2004}]{bri04} Brinchmann J.
et al. 2004, MNRAS, 351, 1151

\bibitem[\protect\citeauthoryear{Bundy et al.}{2006}]{bun06} Bundy K.
et al. 2006, ApJ, 651, 120

\bibitem[\protect\citeauthoryear{Caputi et al.}{2006}]{cap06} Caputi K.I.
et al. 2006, A\&A, 454, 143

\bibitem[\protect\citeauthoryear{Caputi et al.}{2009}]{cap09} Caputi K.I.
et al. 2009, ApJ, 707, 1387

\bibitem[\protect\citeauthoryear{Chang et al.}{2015}]{cha15} Chang Y.,
  van der Wel A., da Cunha E., Rix H. 2015, ApJS, 219, 8

\bibitem[\protect\citeauthoryear{Cid Fernandes et al.}{2005}]{cid05} Cid Fernandes R., 
Mateus A., Sodré L., Stasinska G., \& Gomes J.M. 2005, MNRAS 358, 363

\bibitem[\protect\citeauthoryear{Cimatti, Daddi \& Renzini}{2006}]{cim06}
Cimatti A., Daddi E., Renzini A. 2006, A\&A, 453, 29

\bibitem[\protect\citeauthoryear{Cowie et al.}{1996}]{cow96} Cowie L.L.
et al. 1996, AJ, 112, 839

\bibitem[\protect\citeauthoryear{Crawford, Wirth \& Bershady}{2014}]{cra14}
Crawford S.M., Wirth, G.D., Bershady, M.A. 2014, ApJ, 786, 30

\bibitem[\protect\citeauthoryear{Crossett et al.}{2014}]{cro14} Crossett J.P.
et al. 2014, MNRAS, 437, 2521

\bibitem[\protect\citeauthoryear{da Cunha, Charlot \& Elbaz}{2008}]{dac08}
da Cunha E., Charlot S. \& Elbaz D. 2008, MNRAS, 388, 1595

\bibitem[\protect\citeauthoryear{da Cunha et al.}{2013}]{dac13} da Cunha E.
et al. 2013, ApJ, 765, 9

\bibitem[Djorgovski et al.(2003)]{djo03} Djorgovski 
S.G., de Carvalho R.R., Gal R.R., Odewahn S.C., Mahabal A.A., Brunner R.J.,
Lopes P.A.A., Kohl Moreira J.L. 2003, Bulletin of the Astronomical Society of
Brazil, 23, 197

\bibitem[\protect\citeauthoryear{Dressler}{1980}]{dre80} Dressler A. 1980, ApJ, 236, 351

\bibitem[\protect\citeauthoryear{Dressler \& Gunn}{1983}]{dre83} Dressler A.,
Gunn J.E. 1983, ApJ, 270, 7

\bibitem[\protect\citeauthoryear{Dressler \& Shectman}{1988}]{dre88} Dressler A., 
\& Shectman S.A. 1988, AJ, 95, 985 (DS)

\bibitem[Fadda et al.(1996)]{fad96} Fadda D., Girardi M., Giuricin G.,
et al. 1996, ApJ, 473, 670

\bibitem[\protect\citeauthoryear{Fontanot et al.}{2009}]{fon09} Fontanot F.
et al. 2009, MNRAS, 397, 1776

\bibitem[\protect\citeauthoryear{Gal et al.}{2003}]
{gal03} Gal R.R., de Carvalho R.R., Lopes P.A.A., Djorgovski S.G.,
Brunner R.J., Mahabal A.A., Odewahn S.C. 2003, AJ, 125, 2064

\bibitem[\protect\citeauthoryear{Gal et al.}{2004}]
{gal04} Gal R.R., de Carvalho R.R., Odewahn S.C., Djorgovski S.G., 
Mahabal A.A., Brunner R.J., Lopes P.A.A. 2004, AJ, 128, 3082

\bibitem[\protect\citeauthoryear{Gal et al.}{2009}]
{gal09} Gal R.R., Lopes P.A.A, de Carvalho R.R., Kohl-Moreira J.L.,
Capelato H.V., Djorgovski S.G. 2009, AJ, 137, 2981

\bibitem[\protect\citeauthoryear{Goto}{2005}]{got05} Goto T. 2005,
MNRAS, 357, 937

\bibitem[\protect\citeauthoryear{Guglielmo et al.}{2015}]{gug15} Guglielmo V.
et al. 2015, MNRAS, 450, 2749

\bibitem[\protect\citeauthoryear{Haines et al.}{2007}]{hai07} Haines C.P.
et al. 2007, MNRAS, 381, 7

\bibitem[\protect\citeauthoryear{Haines et al.}{2013}]{hai13} Haines C.P.
et al. 2013, ApJ, 775, 126

\bibitem[\protect\citeauthoryear{Haines et al.}{2015}]{hai15} Haines C.P.
et al. 2015, ApJ, 806, 101

\bibitem[\protect\citeauthoryear{Holden et al.}{2012}]{hol12} Holden B.P.
et al. 2012, ApJ, 749, 96

\bibitem[\protect\citeauthoryear{Kannappan et al.}{2009}]{kan09} Kannappan S.J.
et al. 2009, AJ, 138, 579

\bibitem[\protect\citeauthoryear{Kauffmann et al.}{2004}]{kau04} Kauffmann G., 
White S., Heckman T. et al. 2004, MNRAS, 353, 713

\bibitem[\protect\citeauthoryear{La Barbera et al.}{2010}]{lab10} La Barbera F., 
Lopes P.A.A., de Carvalho R.R., de La Rosa I.G., Berlind A.A. 2010, MNRAS, 408, 1361

\bibitem[\protect\citeauthoryear{Lackner \& Gunn}{2013}]{lac13} Lackner C.N., 
  \& Gunn J.E. 2013, MNRAS, 428, 2141

\bibitem[\protect\citeauthoryear{Lintott et al.}{2008}]{lin08} Lintott C.
  et al. 2008, MNRAS, 389, 1179

\bibitem[\protect\citeauthoryear{Lopes et al.}{2004}]{lop04} Lopes
  P.A.A., de Carvalho R.R., Gal R.R., Djorgovski S.G.,
  Odewahn S.C., Mahabal A.A., Brunner R.J. 2004, AJ, 128, 1017

\bibitem[\protect\citeauthoryear{Lopes et al.}{2006}]{lop06} Lopes
  P.A.A., de Carvalho R.R., Capelato H.V., Gal R.R., Djorgovski S.G.,
  Brunner R.J., Odewahn S.C., Mahabal A.A. 2006, ApJ, 648, 209

\bibitem[\protect\citeauthoryear{Lopes}{2007}]{lop07} Lopes P.A.A. 2007,
MNRAS, 380, 1680

\bibitem[\protect\citeauthoryear{Lopes et al.}{2009a}]{lop09a} Lopes
  P.A.A., de Carvalho R.R., Kohl-Moreira J.L., Jones C. 2009a, MNRAS, 392, 
135, paper I

\bibitem[\protect\citeauthoryear{Lopes et al.}{2009b}]{lop09b} Lopes
  P.A.A., de Carvalho R.R., Kohl-Moreira J.L., Jones C. 2009b, MNRAS, 399, 
2201, paper II

\bibitem[\protect\citeauthoryear{Lopes, Ribeiro \& Rembold}{2014}]{lop14} Lopes
  P.A.A., Ribeiro A.L.B., Rembold S.B., 2014, MNRAS, 437, 2430, paper IV

\bibitem[\protect\citeauthoryear{Maier et al.}{2016}]{mai16} Maier C.
  et al. 2016, arXiv:1602.00686

\bibitem[\protect\citeauthoryear{Masters et al.}{2010}]{mas10} Masters K.
  et al. 2010, MNRAS, 404, 792
  
\bibitem[\protect\citeauthoryear{McIntosh et al.}{2014}]{mci14} McIntosh D.H.
et al. 2014, MNRAS, 442, 533

\bibitem[\protect\citeauthoryear{Menanteau et al.}{2013}]{men06} Menanteau F.
et al. 2006, AJ, 131, 208

\bibitem[\protect\citeauthoryear{Muldrew et al.}{2012}]{mul12} Muldrew S., Croton D., 
Skibba R. et al. 2012, MNRAS, 419, 2670

\bibitem[\protect\citeauthoryear{Muzzin et al.}{2014}]{muz14} Muzzin A.
et al. 2014, ApJ, 796, 65

\bibitem[\protect\citeauthoryear{Noble et al.}{2013}]{nob13} Noble A.G.
et al. 2013, ApJ, 768, 118

\bibitem[\protect\citeauthoryear{Noble et al.}{2016}]{nob16} Noble A.G.
et al. 2016, ApJ, 816, 48

\bibitem[\protect\citeauthoryear{Odewahn et al.}{2004}]
{ode04} Odewahn S.C., de Carvalho R.R., Gal R.R., Djorgovski S.G., 
Brunner R.J., Mahabal A.A., Lopes P.A.A., Kohl Moreira J.L., 
Stalder B. 2004, AJ, 128, 3092

\bibitem[\protect\citeauthoryear{Oemler}{1974}]{oem74} Oemler A. 1974, ApJ, 194, 1

\bibitem[\protect\citeauthoryear{Peng et al.}{2015}]{pen15} Peng Y.,
Maiolino R., Cochrane R. et al. 2015, Nature, 521, 192
  
\bibitem[\protect\citeauthoryear{Ribeiro et al.}{2013}]{rib13} Ribeiro A.L.B.,
  Lopes P.A.A. \& Rembold S.B. 2013, A\&A, 556, 74, paper III

\bibitem[Rines \& Diaferio(2006)]{rin06} Rines K. \& Diaferio A. 2006, ApJ, 
132, 1275

\bibitem[\protect\citeauthoryear{Roberts et al.}{2016}]{rob16} Roberts I.D.,
  Parker L.C., Karunakaran A. 2016, MNRAS, 455, 3628

\bibitem[\protect\citeauthoryear{Salim et al.}{2007}]{sal07} Salim S.
et al. 2007, ApJS, 173, 267

\bibitem[\protect\citeauthoryear{Salim et al.}{2009}]{sal09} Salim S.
et al. 2009, ApJ, 700, 161

\bibitem[\protect\citeauthoryear{Salim \& Rich}{2010}]{sal10} Salim S.
\& Rich R.M. 2010, ApJ, 714, 290

\bibitem[\protect\citeauthoryear{Smethurst et al.}{2015}]{sme15} Smethurst R.J.
et al. 2015, MNRAS, 450, 435

\bibitem[\protect\citeauthoryear{Soifer et al.}{1984}]{soi84} Soifer B.T.
et al. 1984, ApJ, 278L, 71

\bibitem[\protect\citeauthoryear{Strateva et al.}{2001}]{str01}  Strateva I., 
Ivezi\'c Z., Knapp G. et al. 2001, AJ, 122, 1861

\bibitem[\protect\citeauthoryear{Strauss et al.}{2002}]{str02}  Strauss M., Weinberg D., 
Lupton R. et al. 2002, AJ, 124, 1810

\bibitem[\protect\citeauthoryear{Swinbank et al.}{2012}]{swi12} Swinbank A.M.
et al. 2012, MNRAS, 420, 672

\bibitem[\protect\citeauthoryear{Tojeiro et al.}{2013}]{toj13} Tojeiro R.
et al. 2013, MNRAS, 432, 359

\bibitem[\protect\citeauthoryear{Tortora et al.}{2010}]{tor10} Tortora C.
et al. 2010, MNRAS, 407, 144

\bibitem[\protect\citeauthoryear{Tran et al.}{2003}]{tra03} Tran K.H.
et al. 2003, ApJ, 599, 865

\bibitem[\protect\citeauthoryear{Valentinuzzi et al.}{2011}]{val11} Valentinuzzi T.
Poggianti B., Fasano G. et al. 2011, A\&A, 536, 34

\bibitem[\protect\citeauthoryear{Vulcani et al.}{2015}]{vul15} Vulcani B.
et al. 2015, ApJ, 798, 52

\bibitem[\protect\citeauthoryear{Wei et al.}{2010}]{wei10} Wei L.H.
et al. 2010, ApJ, 725L, 62

\bibitem[\protect\citeauthoryear{Wolf et al.}{2009}]{wol09} Wolf C.
et al. 2009, MNRAS, 393, 1302

\bibitem[\protect\citeauthoryear{Wuyts et al.}{2007}]{wuy07} Wuyts S.
et al. 2007, ApJ, 655, 51

\bibitem[\protect\citeauthoryear{Wyder et al.}{2007}]{wyd07} Wyder T.K.
et al. 2007, ApJS, 173, 293

\bibitem[\protect\citeauthoryear{Yee \& L\'opez-Cruz}{1999}]{yee99} Yee H. \& 
L\'opez-Cruz O., 1999, AJ, 117, 1985

\bibitem[\protect\citeauthoryear{Zabludoff et al.}{1996}]{zab96} Zabludoff A.I.
et al. 1996, ApJ, 466, 104


\end{thebibliography}


%%%%%%%%%%%%%%%%%%%%%%%%%%%%%%%%%%%%%%%%%%%%%%%%%%

%%%%%%%%%%%%%%%%% APPENDICES %%%%%%%%%%%%%%%%%%%%%

%\appendix


%%%%%%%%%%%%%%%%%%%%%%%%%%%%%%%%%%%%%%%%%%%%%%%%%%

% Don't change these lines
\bsp	% typesetting comment
\label{lastpage}
\end{document}